%% file: _main.tex
\documentclass[11pt]{article}
\usepackage[dvips]{graphicx}
\usepackage{enumitem}
\usepackage{amssymb}
\usepackage{color}
\usepackage{amsmath}
\usepackage{nicefrac}
\usepackage[left]{lineno}
\usepackage{hyperref}
\usepackage{adjustbox}
\usepackage{xspace}
\usepackage{multirow}
\usepackage[dvipsnames]{xcolor}

\definecolor{blu}{rgb}{0.,0.,1.}
\definecolor{red}{rgb}{1.,0.,0.}
\definecolor{burgundy}{rgb}{0.5, 0.0, 0.13}
\definecolor{crimsonred}{rgb}{0.6, 0.0, 0.0}
\definecolor{persianblue}{rgb}{0.11, 0.22, 0.73}
\definecolor{forestgreen}{rgb}{0.13,0.35,0.13}

\hypersetup{colorlinks, citecolor=crimsonred, linkcolor=persianblue, urlcolor=crimsonred}

\def\authorlist#1#2{
    \vskip 0.4in
\begin{center}\begin{large} {\bf  #1 } \end{large}
    \vskip 0.2in
              #2
     \vskip 0.2in
   \end{center}
}

\newcommand{\chapter}[1]
{
\begin{center}
{\LARGE\bf{#1}}\\
\vspace{10mm}
{\large\bf Report of Topical Group RF2 to Snowmass 2021\\
\vspace{5mm}
Rare Processes and Precision Measurements Frontier
\vspace{-5mm}
}
\end{center}
}

\begin{document}

\newcounter{chapter}

\input lightquarks.tex


\end{document}

%% file: lightquarks.tex
\setcounter{chapter}{1} 



\chapter{Weak Decays of Strange and Light Quarks}

\vspace{-6mm}

\authorlist{\large RF2 conveners:\\ \vspace{1mm} Evgueni Goudzovski$^{1}$, Emilie Passemar$^{2,3}$}
{\vspace{-3mm}\large\bf Community contributors:\\ \vspace{1mm} 
\mbox{Jason Aebischer$^4$},
\mbox{Swagato Banerjee$^5$},
\mbox{Douglas Bryman$^{6,7}$},
\mbox{Andrzej Buras$^8$},
\mbox{Vincenzo Cirigliano$^9$},
\mbox{Norman Christ$^{10}$},
\mbox{Avital Dery$^{11}$},
\mbox{Francesco Dettori$^{12}$},
\mbox{Liping Gan$^{13}$},
\mbox{Corrado Gatto$^{14,15}$},
\mbox{Sergi Gonzalez-Solis$^{16}$},
\mbox{David Hertzog$^{9}$},
\mbox{Luchang Jin$^{17}$},
\mbox{Jacky Kumar$^8$},
\mbox{Andrzej Kupsc$^{18,19}$},
\mbox{Cristina Lazzeroni$^1$},
\mbox{Karim Massri$^{20}$},
\mbox{Matthew Moulson$^{21}$},
\mbox{Hajime Nanjo$^{22}$},
\mbox{Tadashi Nomura$^{23}$},
\mbox{Xin Qian$^{24}$},
\mbox{Diego Redigolo$^{25}$},
\mbox{Giuseppe Ruggiero$^{25}$},
\mbox{Diego Martinez Santos$^{26}$},
\mbox{Chien-Yeah Seng$^{27}$},
\mbox{Vladimir Tishchenko$^{24}$},
\mbox{Kohsaku Tobioka$^{23,28}$},
\mbox{Yau Wah$^{29}$},
\mbox{Elisabeth Worcester$^{24}$},
\mbox{Taku Yamanaka$^{22}$},
\mbox{Jure Zupan$^{30}$}}
\vspace{-5mm}
\footnotesize{
{$^1$School of Physics and Astronomy, University of Birmingham, B15 2TT, United Kingdom\\
$^2$Department of Physics, Indiana University, Bloomington, IN 47405, USA\\
$^3$Departament de F\'isica Te\`orica, IFIC, Universitat de Val\`encia, E-46071 Val\`encia, Spain \\
$^4$Physik-Institut, Universit\"at Z\"urich, CH-8057 Z\"urich, Switzerland\\
$^5$University of Louisville, Louisville, KY 40292, USA\\
$^6$Department of Physics and Astronomy, University of British Columbia, Vancouver, V6T 1Z1, Canada\\
$^7$TRIUMF, 4004 Wesbrook Mall, Vancouver, V6T 2A3, Canada\\
$^8$TUM Institute for Advanced Study, Lichtenbergstr. 2a, D-85747 Garching, Germany\\
$^9$University of Washington, Seattle, WA 98195, USA\\
$^{10}$Physics Department, Columbia University, New York, NY 10027, USA\\
$^{11}$Department of Physics, LEPP, Cornell University, Ithaca, NY 14853, USA\\
$^{12}$INFN Sezione di Cagliari, 09042 Monserrato, Italy\\
$^{13}$University of North Carolina, Wilmington, NC 28403, USA\\
$^{14}$INFN Sezione di Napoli, 80126 Napoli, Italy\\
$^{15}$Northern Illinois University, DeKalb, IL 60115, USA\\
$^{16}$Theoretical Division, Los Alamos National Laboratory, Los Alamos, NM 87545, USA\\
$^{17}$Physics Department, University of Connecticut, Storrs, CT 06269-3046, USA\\
$^{18}$National Centre for Nuclear Research, Warsaw, Poland\\
$^{19}$Uppsala University, 75120 Uppsala, Sweden\\
$^{20}$CERN, CH-1211 Geneva 23, Switzerland\\
$^{21}$INFN Laboratori Nazionali di Frascati, 00044 Frascati, Italy\\
$^{22}$Department of Physics, Osaka University, Toyonaka, Osaka, 560-0043, Japan\\
$^{23}$High Energy Accelerator Research Organization (KEK), Tsukuba 305-0801, Japan\\
$^{24}$Physics Department, Brookhaven National Laboratory, Upton, NY 11973, USA\\
$^{25}$Universit\`a and INFN Sezione di Firenze, 59100 Sesto F.no, Italy\\
$^{26}$IGFAE, Universidade de Santiago de Compostela, 15782 Santiago de Compostela, Spain\\
$^{27}$HISKP and Bethe Center for Theoretical Physics, 
Universit\"{a}t Bonn, 53115 Bonn, Germany\\
$^{28}$Department of Physics, Florida State University, Tallahassee, FL 32306, USA\\
$^{29}$The University of Chicago,
5801 S Ellis Ave, Chicago, IL 60637, USA\\
$^{30}$Department of Physics, University of Cincinnati, Cincinnati, OH 45221, USA\\
}}

\normalsize{}



\newpage
\input rf2_intro.tex
\newpage
\input rf2_kaons.tex
\input rf2_hyperons.tex
\input rf2_pions.tex
\input rf2_eta.tex
\input rf2_summary.tex


\newpage

\bibliographystyle{JHEP}
\bibliography{myreferences}

%% file: rf2_intro.tex
\section{Introduction}
\label{sec:rf2:intro}

An essential feature of flavor physics experiments is their ability to probe very high mass scales, beyond the energies accessible directly in collider experiments, due to quantum effects allowing virtual particles to modify the results of precision measurements in ways that reveal the underlying physics. Therefore the ongoing and planned flavor physics experiments will provide essential constraints and complementary information on the structure of models put forward to explain any discoveries at the LHC and future colliders.
Furthermore, flavor physics experiments are becoming increasingly important laboratories for high-sensitivity searches for feebly interacting hidden sectors.

This report describes the physics case for the studies of weak decays of strange and light quarks. Ongoing and proposed precision measurements of kaon, hyperon, pion and $\eta^{(\prime)}$ meson decays allow for unique tests of the Standard Model (SM). This includes precision measurements of the elements of the CKM quark-mixing matrix leading to stringent unitarity tests; precision symmetry tests including lepton flavor and lepton number conservation; and precision lepton flavor universality tests. In the context of models beyond the SM description (BSM models), strange and light quark decay experiments provide sensitivity to new physics up to the PeV mass scale, as well as leading sensitivities to scenarios involving feebly interacting hidden sectors below the GeV mass scale.

Experimental studies of weak decays of strange and light quarks currently represent a very active field, and significant progress is expected over the next decade. Two dedicated high-intensity kaon experiments focusing on unique SM tests with ultra-rare decays, NA62 and KOTO, are currently taking data. Moreover, both collaborations are developing ambitious long-term programs. The LHCb experiment at CERN is pursuing a kaon and hyperon decay program, while the BESIII experiment is conducting pioneering searches for direct CP violation in hyperon decays. A new major initiative focused on a precision lepton universality test and a measurement of the mixing matrix element $V_{ud}$ in charged pion decays, PIONEER, has been approved recently at the Paul Scherrer Institute. 
The Belle~II experiment will provide important complementary information from $\tau$ decays, including a $V_{us}$ measurement.
The JEF $\eta$ factory under construction at the Jefferson Lab will focus on symmetry tests and searches for hidden sectors in rare $\eta^{(\prime)}$ decays, and a next-generation $\eta^{(\prime)}$ factory proposal, REDTOP, has been put forward. Theory and lattice QCD are making important steps essential for the interpretation of measurements.

The US should take advantage of the above medium-scale initiatives, many of which are centered in Europe and Asia. These experiments lead to powerful physics insights on relatively short time scales, offer opportunities of making leading contributions to potential BSM discoveries, and provide comprehensive experimental training at various stages including experimental design, R\&D, detector construction, data collection and analysis.

%% file: rf2_kaons.tex
\section{Kaon decays}
\label{sec:rf2:kaons}

\subsection{Kaon decays in the Standard Model and beyond}
\label{sec:rf2:kaons:theory}

Kaon decay studies have played a unique role in the establishment of the Standard Model of particle physics over the past 70~years: CP violation, the suppression of flavor changing neutral currents (FCNC) and the GIM mechanism have been discovered in the kaon system. Because of the relatively small number of decay modes, simple final states, and availability of high-intensity kaon beams leading to large datasets of ${\cal O}(10^{13})$ decays, kaon decay experiments continue to be in many ways the quintessential intensity-frontier experiments. At present, kaon physics is focused on precision measurements of highly suppressed loop-induced FCNC processes that may reveal the effects of BSM physics at the 100~TeV scale, above the scale that can be explored directly by the LHC or even a next-generation hadron collider~\cite{Aebischer:2022vky,Buras:2014zga}. Kaon decay measurements are complementary to those performed in the $B$ sector, in particular by providing crucial information for the interpretation of the flavor anomalies observed in $B$ decays. Progress in kaon decay physics implies a large amount of experimental work on precision measurements at high-luminosity, using dedicated detectors based on state-of-the-art technologies. On the other hand, significant theoretical work is required to understand the correlations with the other sectors of flavor physics, and to quantify the low-energy effects hindering the interpretation of the measurements.


The well-established rare kaon decays of interest include those for which the short-distance (SD) amplitudes are CP-conserving ($K_L\to\mu^+\mu^-$, $K_S\to\pi^0\ell^+\ell^-$), and those with the SD amplitudes receiving CP-violating contributions in the SM and many BSM scenarios ($K^+\to\pi^+\nu\bar\nu$, $K_L\to\pi^0\nu\bar\nu$, $K_S\to\mu^+\mu^-$, $K_L\to\pi^0\ell^+\ell^-$), thereby opening a possibility of searching for new sources of CP violation~\cite{Aebischer:2022vky}. The rates of these decays are both extremely suppressed and accurately predicted within the SM. Moreover it has been pointed out recently that $K_S\to\mu^+\mu^-$ is another ``golden mode'' providing a complementary, theoretically clean observable~\cite{DAmbrosio:2017klp,Dery:2021mct}. While the total $K\to\mu^+\mu^-$ rate is dominated by long-distance (LD) physics, the measurement (in a dedicated high-intensity $K_S$ experiment) of CP violation in the interference of mixing and decay via the time-dependent rate enables the extraction of the purely CP odd short-distance amplitude, which is predicted within the SM with an ${\cal O}(1\%)$ uncertainty. Beyond ultra-rare FCNC processes, studies of rare kaon decays provide insight into low-energy QCD, and inputs for the interpretation of the flagship measurements~\cite{Cirigliano:2011ny}.

The current experimental efforts are focused on the $K\to\pi\nu\bar\nu$ decays which have a uniquely clean theoretical character due to the small LD contributions. In the standard approach~\cite{Buras:2015qea}, the main theoretical uncertainties in the SM prediction for the $K\to\pi\nu\bar\nu$ branching ratios stem from the CKM parameters ($V_{cb}$, $V_{ub}$ and the $\gamma$ angle), leading to a 10\% accuracy. A novel approach, proposed recently to eliminate the dependence on $|V_{cb}|$ and $\gamma$, leads to a 5\% precision~\cite{Buras:2021nns,Buras:2022wpw,Buras:2022qip}:
\begin{eqnarray}
{\cal B}_{\rm SM}(K^+\to\pi^+\nu\bar\nu) & = & (8.60\pm0.42)\times 10^{-11}, \nonumber \\
{\cal B}_{\rm SM}(K_L\to\pi^0\nu\bar\nu) & = & (2.94\pm0.15)\times 10^{-11}. \nonumber
\end{eqnarray}
The non-parametric uncertainties above are 4\% (dominated by a LD charm contribution) in the $K^+$ case, and 1.5\% in the $K_L$ case. The accuracy of the SM predictions is expected to improve to about 3\% over the next decade, due to lattice QCD progress on the charm contribution~\cite{Bai:2018hqu,Christ:2019dxu,Blum:2022wsz} and reduction of the external parametric uncertainties. Further advances due to perturbative calculations are unlikely.

\begin{figure}[t]
\begin{center}
\includegraphics[width=0.6\textwidth]{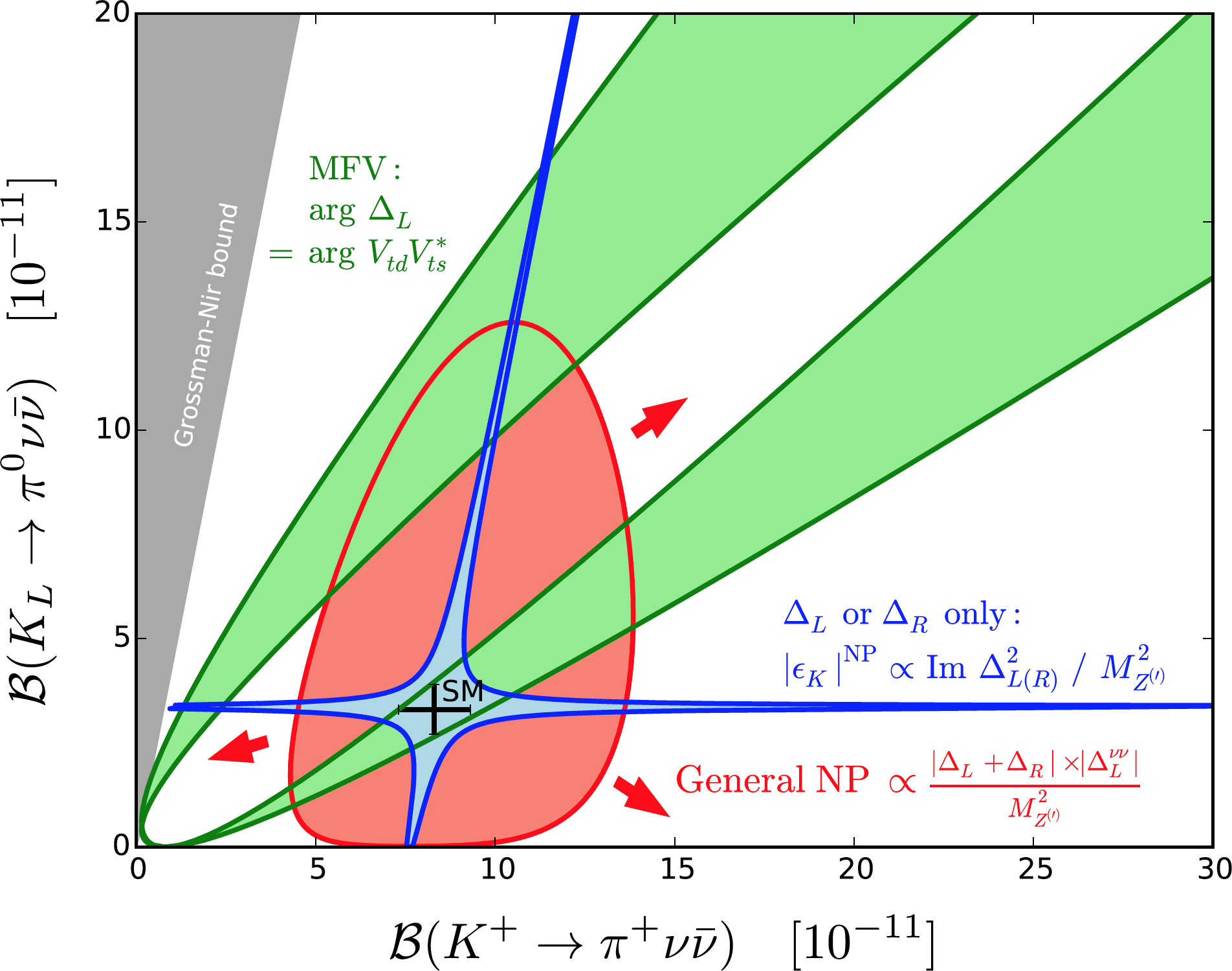}
\end{center}
\vspace{-12mm}
\caption{Illustration of the expected modifications of $K\to\pi\nu\bar\nu$ branching ratios in several generic BSM scenarios~\cite{Buras:2015yca}. Note that precision measurement of both decay modes allows determination of the $\beta$ angle within the SM paradigm~\cite{Buchalla:1994tr}.}
\label{fig:KpnnBSM}
\end{figure}

The $K\to\pi\nu\bar\nu$ decays have been analysed thoroughly in numerous SM extensions~\cite{Aebischer:2022vky}. The general scheme for the expected correlation between the $K^+$ and $K_L$ decays in various scenarios is illustrated in Fig.~\ref{fig:KpnnBSM} reproduced from Ref.~\cite{Buras:2015yca}. Due to the different sensitivity of the two decays to new sources of CP violation, measurements of both $K^+\to\pi^+\nu\bar\nu$ and $K_L\to\pi^0\nu\bar\nu$ modes are crucial to uncover the possible evidence for new physics in the quark flavor sector, and to distinguish between various classes of new physics
models.

More generally, BSM contributions to $K\to\pi\nu\bar\nu$ decays are expected to be correlated with BSM effects in other observables. Important correlations have been established with the CP violation parameter $\varepsilon'/\varepsilon$ and the CP conserving $K_L-K_S$ mass difference $\Delta M_K$ in the kaon sector~\cite{Buras:2015yca,Buras:2015jaq,Aebischer:2020mkv}, the rates of rare $B$ meson decays $B\to K\nu\bar\nu$ and $B\to\mu^+\mu^-$~\cite{Altmannshofer:2009ma,Buras:2014fpa}, and the possible lepton flavor universality violating effects in $B$ decays~\cite{Bordone:2017lsy}. Furthermore, experimental study of the $K^+\to\pi^+\nu\bar\nu$ kinematic distribution allows testing the Dirac vs Majorana nature of the neutrino, probing ${\cal O}(10~{\rm TeV})$ operator scales in different quark and neutrino flavors compared to neutrinoless double beta decay~\cite{Deppisch:2020oyx}.

In spite of their significant sensitivity to new physics, no new experiments are currently planned to measure with increased precision direct or indirect CP violation in $K^0\to\pi\pi$ decays, or the mass difference $\Delta M_K$, since the current experimental precision on these quantities exceeds the theoretical uncertainties in their SM prediction. Nevertheless the quantities $\varepsilon$, $\varepsilon'$ and $\Delta M_K$ offer opportunities for the discovery of new physics as the improving capabilities of lattice QCD provide increasingly accurate SM predictions~\cite{Christ:2015phf, RBC:2020kdj, Wang:2021twm}. Over the next decade, the accuracy of the SM predictions for $\Delta M_K$, $\varepsilon$ and $\varepsilon'$ is expected to improve by a factor of five in each case. If this is achieved for the calculation of $\varepsilon'$, the theoretical precision would match that of the experiment, $\textrm{Re}(\varepsilon'/\varepsilon) = 16.6(2.3) \times 10^{-4}$~\cite{NA48:2002tmj,KTeV:2010sng}, potentially motivating a new measurement.

\begin{figure}[p]
\begin{center}
\includegraphics[width=0.57\textwidth]{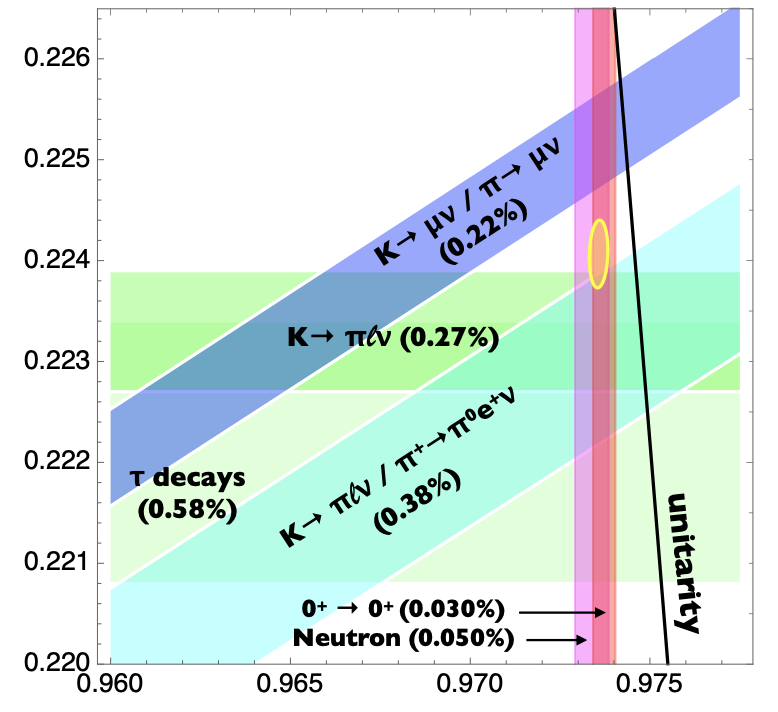}
\put(-280,205){\rotatebox{90}{\boldmath\LARGE $V_{us}$}}
\put(-48,-10){\boldmath\LARGE $V_{ud}$}
\end{center}
\vspace{-11mm}
\caption{Constraints on $V_{ud}$ and $V_{us}$ obtained from measurements of kaon, pion, $\tau$ lepton, nuclear and neutron decays~\cite{Bryman:2021teu}, leading to the Cabibbo angle anomaly.} 
\label{fig:VusVud}
\end{figure}

Measurements of semileptonic kaon decays $K\to\pi\ell\nu$ provide the principal input for the extraction of the CKM parameter $V_{us}$ while the ratios of (semi)leptonic $K^+$ and $\pi^+$ decay rates are used to extract the ratio $V_{us}/V_{ud}$, with inputs provided from lattice QCD~\cite{Aoki:2021kgd}. Determination of $V_{us}$ from kaon, pion and $\tau$ decays, combined with $V_{ud}$ measurement from super-allowed beta decays~\cite{Hardy:2020qwl} and neutron decays~\cite{Czarnecki:2018okw,Seng:2020wjq}, gives rise to a $3\sigma$ deficit in first-row CKM unitarity relation; a tension of similar significance is observed between $K\to\ell\nu$ and $K\to\pi\ell\nu$ rates~\cite{Cirigliano:2022yyo,Bryman:2021teu}. The resulting Cabibbo angle anomaly is illustrated, and quantified, in Fig.~\ref{fig:VusVud}.%
\footnote{Note that $N_f = 2+1+1$ FLAG average for $f_+(0)$~\cite{Aoki:2021kgd} is used. A recent result for $N_f = 2+1$ leads to a smaller $f_+(0)$ value~\cite{Ishikawa:2022otj} while compatible within 1$\sigma$ with the $N_f=2+1$ FLAG average, and within 1.5$\sigma$ with the $N_f=2+1+1$ value.}
The uncertainty in $V_{us}$ comes in equal parts from the experimental errors and theoretical uncertainties in the ratio of decay constants, $f_K/f_\pi$, and the $K\to\pi\ell\nu$ form factor, $f_+(0)$. Substantial improvements in the lattice QCD calculations of these hadronic factors are expected in the next five years, thanks to decreased lattice spacing and accurate evaluation of
electromagnetic effects~\cite{Carrasco:2015xwa, Boyle:2019rdx,Feng:2021zek,Seng:2020jtz}. Note that significant progress on the calculation of radiative corrections
has been achieved recently, reducing the electromagnetic correction uncertainties to a negligible level~\cite{Seng:2019lxf,Seng:2021boy,Seng:2021wcf,Seng:2022wcw}.
It is important that improvements in the accuracy of kaon decay measurements are achieved; in particular a precision measurement of the ratio of $K^+\to\pi^0\mu^+\nu$ and $K^+\to\mu^+\nu$ rates is well-motivated~\cite{Cirigliano:2022yyo}.

It should be noted that the Belle~II experiment is planning to extract $V_{us}$ at an improved precision, with respect to the $\tau$-decay band shown in Fig.~\ref{fig:VusVud}, from a dedicated suite of inclusive and exclusive [${\cal B}(\tau^-\to K^-\nu)/{\cal B}(\tau^-\to\pi^-\nu)$] measurements of $\tau$ decays, combined with theory improvements~\cite{Belle-II:2022cgf}. Measurements of inclusive hadronic decays of the $\tau$ lepton provide a unique opportunity to extract $|V_{us}|$ which does not involve theoretically estimated hadronic form factors or decay constants, therefore providing a possibility to cross-check the kaon results. The theory inputs in this case are based on operator product expansion and finite energy sum rules.


\begin{figure}[p]
\begin{center}
\includegraphics[width=0.8\textwidth]{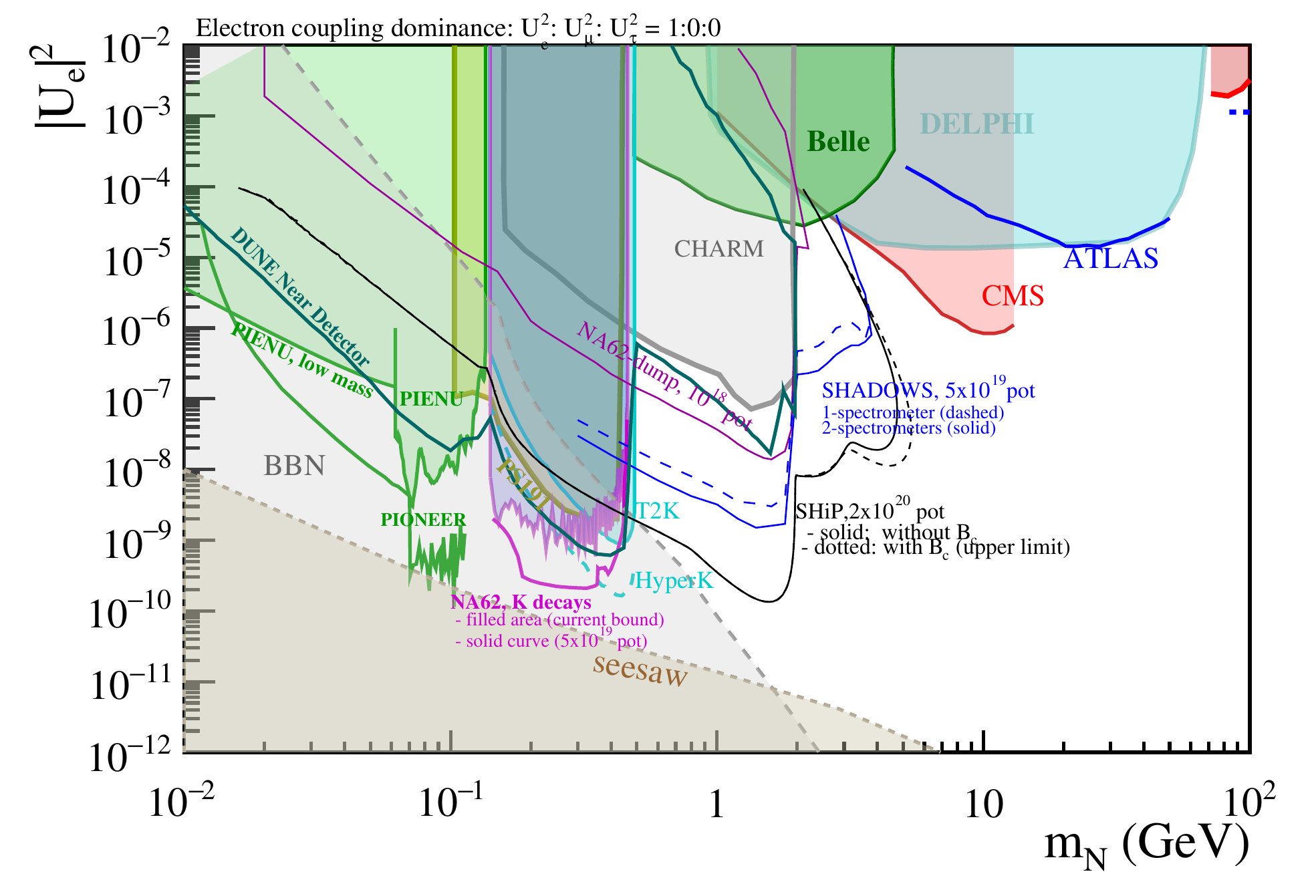}
\end{center}
\vspace{-14mm}
\caption{Existing (solid areas) and projected upper limits on electron couplings of heavy neutral leptons~\cite{Abdullahi:2022jlv}. World-leading limits below 400~MeV/$c^2$ come from $K^+\to e^+N$ searches at NA62~\cite{NA62:2020mcv}, $\pi^+\to e^+N$ searches at PIENU~\cite{PIENU:2017wbj}, and lepton flavor universality tests at PIENU~\cite{Bryman:2019bjg}. Present (NA62 Run~2) and future $K^+$ experiments at CERN, and Phase I of the PIONEER experiment approved at PSI, are expected to bring further improvements towards the seesaw bound.} 
\label{fig:rf2:hnl}
\end{figure}


The kaon sector offers opportunities for precision lepton flavor universality tests by measuring the ratios ${\cal B}(K^+\to\pi^+e^+e^-)/{\cal B}(K^+\to\pi^+\mu^+\mu^-)$, ${\cal B}(K^+\to e^+\nu)/{\cal B}(K^+\to\mu^+\nu)$ and ${\cal B}(K\to\pi e\nu)/{\cal B}(K\to\pi\mu\nu)$, exploiting the cancellation of hadronic effects~\cite{Bryman:2021teu,Crivellin:2016vjc}. Searches for lepton flavor and lepton number violating $K^+\to\pi^-(\pi^0)\ell^+\ell^+$, $K^+\to\pi(\pi^0)\mu e$, $K_L\to(\pi^0)(\pi^0)\mu^\pm e^\mp$ and $K_L\to e^\pm e^\pm \mu^\mp\mu^\mp$ decays, with sensitivities to branching ratios at the ${\cal O}(10^{-12})$ level, represent interest in the context of BSM models involving flavor-violating ALPs~\cite{Cornella:2019uxs} or massive Majorana neutrinos. The constraints on active-sterile mixing angles between Majorana neutrinos obtained from the experimental limits on the $K^+\to\pi^-\ell^+\ell^+$ decays are stronger than those 
obtained from neutrinoless double beta decay searches below the kaon mass~\cite{Abada:2017jjx}.

Kaon decays represent uniquely sensitive probes of light hidden sectors due to the availability of large datasets and the suppression of the kaon decay width. The possible search strategies have been reviewed recently~\cite{Goudzovski:2022vbt}, and the following have been identified as the most promising. Searches for the $K\to\pi X_{\rm inv}$ decay, where $X_{\rm inv}$ represents an invisible particle,
by extension of the $K\to\pi\nu\bar\nu$ measurements represent a unique probe into the Higgs mixed dark scalar and ALP phase space. Searches for resonances in the $K\to\pi\ell^+\ell^-$ and $K\to\pi\gamma\gamma$ decay spectra are complementary to searches at beam-dump experiments for a significant ALP mass range. Searches for heavy neutral lepton ($N$) production in $K^+\to\ell^+N$ decays are approaching the seesaw neutrino mass models with ${\cal O}(100~{\rm MeV})$ sterile neutrinos~\cite{Abdullahi:2022jlv}, as illustrated for the electron coupling in Fig.~\ref{fig:rf2:hnl}. Finally, searches for a leptonic force mediator ($X$) in $K^+\to\mu^+\nu X$ decays can probe a region of parameter space providing an explanation for the muon $g-2$ anomaly~\cite{Krnjaic:2019rsv}. 



\subsection{Current and planned kaon experiments}
\label{sec:rf2:kaons:exp}

The availability of high-intensity kaon beams at the CERN SPS North Area and the J-PARC Hadron Experimental Facility opens unique possibilities for precision SM tests in the quark flavor sector. Significant experimental progress in kaon decay measurements is expected in the coming decade through the exploitation of dedicated operating experiments (NA62 at CERN and KOTO at J-PARC). In the long term, building on the success of the current generation of experiments, ambitious initiatives are taking shape both at CERN and J-PARC to measure the  $K^+\to\pi^+\nu\bar\nu$ branching ratio to 5\% and the $K_L\to\pi^0\nu\bar\nu$ branching ratio to 20\% precision~\cite{KOTO:2022hwn}. These planned experiments would also pursue a comprehensive physics program including lepton flavor universality tests, lepton flavor and number conservation tests, precision rare decay measurements, and searches for hidden sectors.

The main goal of the NA62 experiment at CERN is the measurement of the $K^+\to\pi^+\nu\bar\nu$ decay. The experiment operates a high-intensity unseparated $K^+$ beam with a momentum of 75~GeV/$c$ derived from primary 400~GeV/$c$ protons extracted from the SPS accelerator in spills of 4.8~s duration, and measures kaon decays in a 60~m long fiducial region located in a vacuum tank~\cite{NA62:2017rwk}. The setup includes trackers and Cherenkov detectors for the incoming kaons and for the final-state particles, as well as calorimeters and a hermetic photon veto system. The key detectors deliver a sub-100~ps time precision, thereby providing clean event reconstruction and manageable random veto effects at the nominal 45~MHz beam kaon rate. The signal signature consists of an incoming $K^+$ in time with a final-state $\pi^+$, not kinematically compatible with $K^+\to\mu^+\nu$, $K^+\to\pi^+\pi^0$ and $K^+\to 3\pi$ decays (which defines two signal regions in terms of the missing mass), and no extra in-time activity. 
Background suppression relies on ${\cal O}(10^{-8})$ muon rejection by particle identification, and ${\cal O}(10^{-8})$ rejection of $\pi^0\to\gamma\gamma$ decays by the photon veto system.

\begin{figure}
\begin{center}
\includegraphics[width=0.40\hsize]{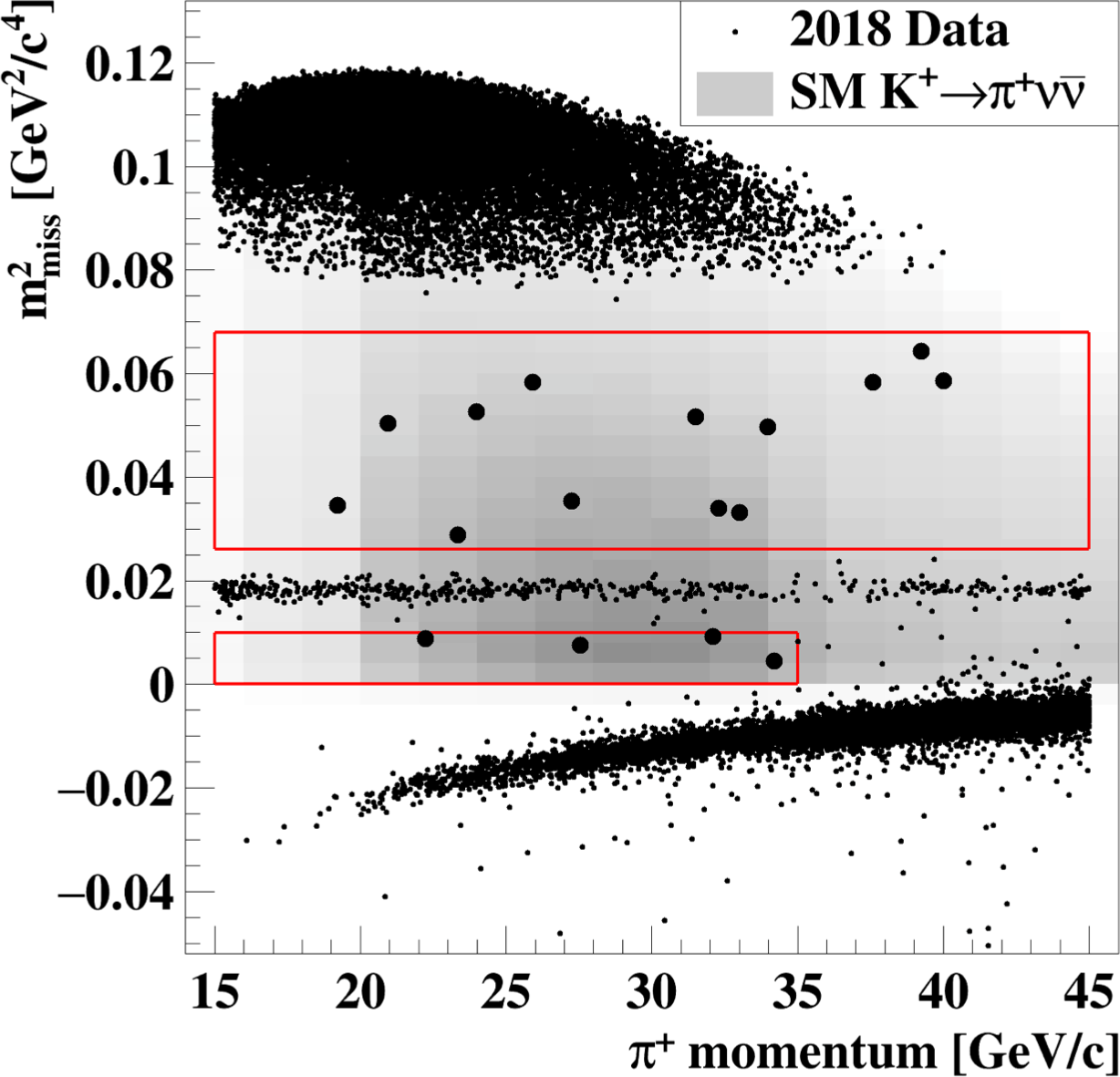}%
\includegraphics[width=0.60\hsize]{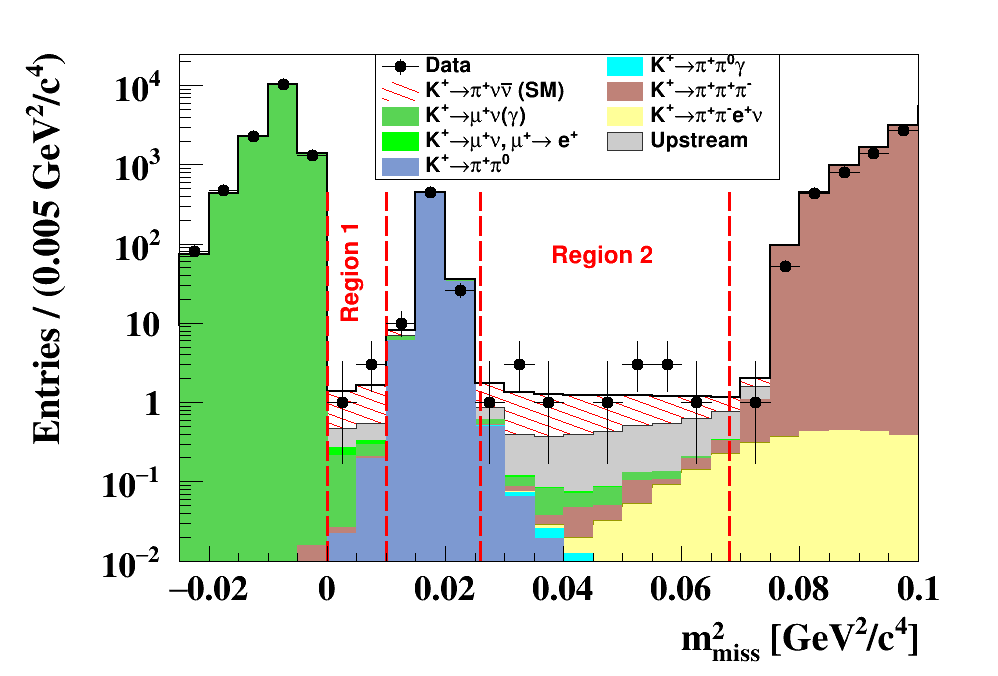}
\end{center}
\vspace{-12mm}
\caption{Distribution of the seventeen $K^+\to\pi^+\nu\bar\nu$ candidates observed by the NA62 experiment in the 2018 dataset~\cite{NA62:2021zjw} in terms of reconstructed $\pi^+$ momentum in the laboratory frame and squared missing mass $m_{\rm miss}^2$ (left), and its $m_{\rm miss}^2$ projection (right). The two signal regions free from the main backgrounds are shown with rectangles in the left panel, and the backgrounds (estimated mainly with data-driven methods) are shown in the right panel.}
\label{fig:na62-run1-Kpinn}
\end{figure}

The NA62 Run~1 dataset, collected in 2016--2018 at a typical beam intensity of $2.2\times 10^{12}$ protons on target (POT) per spill, amounts in total to $2.2\times 10^{18}$ POT and about $6\times 10^{12}$ useful $K^+$ decays. Analysis of this sample has led to an evidence for the $K^+\to\pi^+\nu\bar\nu$ decay with $3.4\sigma$ significance, based on 20~signal candidates with an estimated background of $7.03^{+1.05}_{-0.82}$ events dominated by $K^+$ decays upstream of the decay region~\cite{NA62:2021zjw}. The branching ratio is measured to be \begin{displaymath}
{\cal B}(K^+\to\pi^+\nu\bar\nu) = \left(10.6\,^{+4.0}_{-3.4}\left|_{\rm stat}\pm 0.9_{\rm syst}\right.\right)\times 10^{-11},
\end{displaymath}
and the kinematic distribution of the candidates observed in the 2018 dataset is shown in Fig.~\ref{fig:na62-run1-Kpinn}. This represents an improvement in sensitivity with respect to an earlier result from the BNL E787/E949 experiments, ${\cal B}(K^+\to\pi^+\nu\bar\nu)=(1.73^{+1.15}_{-1.05})\times 10^{-10}$~\cite{BNL-E949:2009dza}.


The experience of NA62 Run~1 has firmly established the decay-in-flight technique. The setup has subsequently been upgraded, including installation of an additional beam tracker station, installation of new veto detectors both upstream and downstream of the decay region, and improvements to the trigger and data acquisition (TDAQ) system~\cite{NA62:2021xxx}. These upgrades have brought the $K^+\to\pi^+\nu\bar\nu$ measurement in a low-background, high-acceptance regime, primarily due to the reduction of background from upstream decays. NA62 Run~2 started in 2021 (reaching the nominal proton beam intensity of $3.3\times 10^{12}$ POT/spill), and is approved until LHC Long Shutdown~3 (LS3), i.e. until the end of 2025, aiming at a 10\% precision on the $K^+\to\pi^+\nu\bar\nu$ rate measurement.

A comprehensive program (HIKE -- ``High-Intensity Kaon Experiments'') for the study of the rare decays of $K^+$ and $K_L$ mesons with high-intensity beams after LS3 is envisioned at CERN, to be carried out in multiple phases using shared detectors and infrastructure~\cite{KOTO:2022hwn}. The program includes an experiment to measure ${\cal B}(K^+\to\pi^+\nu\bar\nu)$ at the 5\% level, an experiment to measure ${\cal B}(K_L\to\pi^0\nu\bar\nu)$ at the 20\% level, and an experiment with a $K_L$ beam and a final-state particle tracker focused measurements of the $K_L\to\pi^0\ell^+\ell^-$ decays~\cite{Mescia:2006jd} while also allowing the characterization of the $K_L$ beam. Collection of up to $10^{19}$ POT in beam-dump mode is also foreseen, providing unique sensitivity to forward production of feebly-interacting particles; beam-dump operation in conjunction with an off-axis detector has been proposed~\cite{Baldini:2021hfw}. These experiments rely on the delivery of a high-intensity proton beam from the SPS to the present NA62 experimental hall ($2.1\times 10^{13}$ POT/spill, $1.3\times 10^{19}$ POT/year). With duty cycle optimization, the required proton flux can be delivered, allowing high-intensity kaon experiments to run as part of a robust fixed-target program in the CERN North Area~\cite{Bartosik:2018xxx}. It should be noted that SPS operation is foreseen until at least 2038.

The planned $K^+\to\pi^+\nu\bar\nu$ experiment at CERN aiming at 400 SM candidates with a signal-to-background ratio $S/B\gg 1$ is based on similar principles to NA62, and is expected to employ a similar experimental layout. The experiment would operate at up to six times the nominal NA62 beam intensity, necessitating a substantial upgrade to bring the time resolutions to the 20~ps level. Design studies for the principal detectors (including a silicon pixel beam tracker, a straw downstream tracker, Cherenkov detectors for particle identification, and the photon veto system) are in progress, and a Letter of Interest is in preparation. The baseline design for the $K_L\to\pi^0\nu\bar\nu$ experiment at CERN is the KLEVER project aiming to collect 60~SM events with $S/B=1$~\cite{Ambrosino:2019qvz}. A high-energy $K_L$ beam, with a mean momentum of 40~GeV/$c$, facilitates the rejection of $K_L\to\pi^0\pi^0$ and other backgrounds by detection of the additional photons in the final state. Extensive design studies performed include the options for the upgrade of proton beam transport~\cite{Banerjee:2018xxx,Gatignon:2018xxx} and secondary neutral beam production and transport~\cite{VanDijk:2019oml} required to reach the design beam intensity, as well as layout optimization and physics performance studies~\cite{Beacham:2019nyx}. Suppression of the $\Lambda$ decay background is likely to require an extension of the experimental hall. Baseline designs of the main electromagnetic calorimeter, small and large-angle photon vetoes and charged-particle rejection detectors have been proposed, and beam tests of the prototypes have been performed.


The main goal of the KOTO experiment at J-PARC is the measurement of the $K_L\to\pi^0\nu\bar\nu$ decay. The experiment derives its $K_L$ beam with a peak momentum of 1.4~GeV/$c$ from the primary 30~GeV protons extracted from the J-PARC main ring. Kaon decays in a 2~m long fiducial region are studied using a detector consisting of a CsI electromagnetic calorimeter and a hermetic system of veto counters surrounding the decay volume. The signal signature consists of two in-time photons from the $\pi^0\to\gamma\gamma$ decay detected in the CsI calorimeter, consistent with emission in the decay volume under the $\pi^0$ mass hypothesis, with a large transverse momentum. Using the 2015 dataset corresponding to $2.2\times 10^{19}$ POT, a single event sensitivity of $1.3\times 10^{-9}$ has been achieved, and an upper limit has been obtained for the decay branching ratio by observing no signal candidates~\cite{KOTO:2018dsc}:
\begin{displaymath}
{\cal B}(K_L\to\pi^0\nu\bar\nu) < 3.0\times 10^{-9} \quad {\rm at~90\%~CL}.
\end{displaymath}

In the 2016--2018 dataset corresponding to $3.05\times 10^{19}$ POT, three signal candidates are observed, which is statistically consistent with the expected background of $1.22\pm0.26$ events dominated by $K^\pm\to\pi^0 e^\pm\nu$ decays of charged kaons produced in charge-exchange reactions of the $K_L$ beam halo in the collimators~\cite{KOTO:2020prk}. The above background estimate is based on the measured beam halo and $K^\pm$ fluxes. While the single event sensitivity of $7.2\times 10^{-10}$ (corresponding to 0.04 SM signal events) improves on the 2015 data, the upper limit obtained on the decay branching ratio is weaker due to the apparent background fluctuation.

The KOTO dataset collected in 2019--2021 is twice the size of the 2016--2018 dataset. An upstream veto detector was installed in 2021 to reduce the charged-kaon background. A further dataset is expected to be collected in 2022--2025, with a gradual increase of the beam intensity (up to 100~kW beam power) and an 
upgraded TDAQ system. The foreseen improvements to the setup include installation of a second upstream veto detector and an upstream sweeping magnet, which is expected to suppress the charged-kaon flux to a negligible level and bring the measurement into a low-background regime. It is expected to reach the SM single event sensitivity of $O(10^{-11})$ by 2025.

The experience gained by KOTO has allowed planning for a next-generation experiment, KOTO step-2~\cite{KOTO:2022hwn}. The proposal relies on the foreseen extension of the J-PARC Hadron Hall, and involves a complete rebuild of the detector. A redesigned beamline would provide a higher $K_L$ momentum (peaking at 3 GeV/$c$, and improved $K_L$ flux, and a reduced halo and neutron flux. The proposed detector involves a 12~m long decay region and a calorimeter with a 3~m diameter, leading to a factor of 5 larger acceptance with respect to KOTO. A sensitivity estimate is reported in Ref.~\cite{KOTO:2022hwn}: the expected SM signal yield is 35 events while the estimated background is $56\pm3$~events (with the largest contribution from $K_L\to\pi^0\pi^0$ decay with two undetected photons), assuming a running time of $3\times 10^7$~s with a 100~kW beam. This is sufficient for a $5\sigma$ discovery of the $K_L\to\pi^0\nu\bar\nu$ decay, assuming the SM branching ratio.

Current and proposed $K\to\pi\nu\bar\nu$ experiments at CERN and J-PARC pursue a broad physics program beyond the flagship measurements (see Section~\ref{sec:rf2:kaons:theory}). As an example, the availability of auxiliary trigger lines during NA62 Run~1 has already allowed lepton flavor and number conservation tests~\cite{NA62:2019eax,NA62:2021zxl,NA62:2022tte},
and searches for hidden-sector particle production in kaon decays~\cite{NA62:2020mcv,NA62:2019meo,NA62:2020xlg,NA62:2020pwi,NA62:2021bji}. Similarly, KOTO has published several searches for forbidden decays and hidden-sector mediators~\cite{KOTO:2018dsc,KOTO:2020bhx}. The sensitivities of the future experiments to the branching fractions of the forbidden kaon decays are expected reach ${\cal O}(10^{-12})$. It should also be noted that kaon experiments serve as ``$\pi^0$ factories'' allowing for precision measurements of rare decays of tagged, Lorentz-boosted $\pi^0$ mesons.

Considering the recent hints for lepton flavor universality violation~\cite{Fischer:2021sqw}, lepton universality tests in kaon decays at the present and future $K^+\to\pi^+\nu\bar\nu$ experiments at CERN deserve a specific discussion. The ratio ${\cal B}(K^+\to\pi^+e^+e^-)/{\cal B}(K^+\to\pi^+\mu^+\mu^-)$ is expected to be measured to a 1\% accuracy from the first dedicated analysis with reduced intrinsic and external uncertainties, improving by a factor of about five on the present status~\cite{ParticleDataGroup:2020ssz}. A measurement of the ratio $R_{\mu e}^{K}={\cal B}(K^+\to e^+\nu)/{\cal B}(K^+\to\mu^+\nu)$ at the per mil level is feasible, improving on the present precision of $R_{\mu e}^{K{\rm exp}}=2.488(9)\times 10^{-5}$~\cite{ParticleDataGroup:2020ssz} dominated by a result from an early phase of the NA62 experiment carried out with the NA48/2 detector~\cite{NA62:2012lny}. However a dedicated, optimised experiment is probably required to match accuracy of the SM prediction $R_{\mu e}^{K{\rm SM}}=2.477(1)\times 10^{-5}$~\cite{Cirigliano:2007xi}, in terms of both statistical and systematic uncertainties. 

The LHCb experiment at the LHC, optimized for short-lived beauty hadrons in terms of the decay length and the transverse momenta of the decay products, carries out a complementary program of $K_S$ decay measurements exploiting the ${\cal O}(1~{\rm barn})$ cross-section for production of kaons in the experiment’s acceptance. With a total reconstruction and trigger efficiency for $K_S$ and hyperon decays of ${\cal O} (10^{-4})$ for Run~1 and Run~2 data, LHCb has provided several world-leading results, including an upper limit ${\cal B}(K_S\to\mu^+\mu^-) < 2.1\times 10^{-10}$ at 90\% CL obtained with the full dataset collected to date~\cite{LHCb:2020ycd}. An upgraded trigger to be operated during Run~3 is expected to bring significant improvements in sensitivity~\cite{AlvesJunior:2018ldo}.

In summary, kaon decays are highly sensitive probes of the flavor and CP violating sector of any SM extension. 
The ambitious initiatives for next-generation kaon experiments at CERN and J-PARC represent a significant increase in scope relative to existing experiments. While the US community is exploring possible expansions to the physics program that could be achieved with future upgrades to the Fermilab proton accelerator complex, participation in kaon experiments at CERN and/or J-PARC is currently the only opportunity for US physicists to contribute to this vital area of research.

%% file: rf2_hyperons.tex
\section{Hyperon decays}
\label{sec:rf2:hyperons}

Studies of hyperon decays are complementary to kaon physics, as their baryon number and spin properties provide different sensitivity to BSM interactions.
While CP violation in kaon decays was discovered back in 1964, it is yet to be established in hyperon decays. Important efforts in this direction were made in the first decade of the century by the HyperCP experiment at Fermilab which studied decays of hyperons produced by a 800~GeV/$c$ proton beam incident on a target~\cite{HyperCP:2004kbv,HyperCP:2004zvh}.

A novel approach is employed by the BESIII experiment, where the copious production of spin-entangled hyperon-antihyperon pairs at the $J/\psi$ resonance is exploited to make direct comparison of the baryon and antibaryon decay properties. A recent proof-of-concept measurement of the direct CP violating asymmetry of the $\Lambda$ and $\bar\Lambda$ decay parameters to a 0.5\% accuracy, based on a sample of $10^{10}$ $J/\psi$ events collected in 2017--2019, represents the most precise CP symmetry test in the hyperon sector to date~\cite{BESIII:2022qax}. Further measurements of weak phases and CP violating asymmetries performed by BESIII include those in the $J/\psi\to\Sigma^+\bar\Sigma^-$~\cite{BESIII:2020fqg} and $J/\psi\to\Xi^-\bar\Xi^+$~\cite{BESIII:2021ypr} decays; no asymmetry is observed in either case. A next-generation $J/\psi$ factory is expected to collect a dataset of ${\cal O}(10^{12})$ $J/\psi$ events, and the statistical precision of the CP tests can be improved using longitudinally-polarized electron beams~\cite{Salone:2022lpt}. This would allow observation of statistically significant asymmetries at the ${\cal O}(10^{-4})$ level of the SM prediction~\cite{Tandean:2002vy}, providing further insight into the mechanism of CP violation in the baryon sector.

The BESIII experiment is also pursuing a program of rare hyperon decay measurements~\cite{BESIII:2021ynj,BESIII:2021emv}. In particular, upper limits have recently been established on the rates of lepton number violating hyperon decays: $\mathcal{B}(\Sigma^{-} \to p e^{-} e^{-}) < 6.7 \times 10^{-5}$ and $\mathcal{B}(\Sigma^{-} \to \Sigma^{+} X) < 1.2 \times 10^{-4}$ at 90\% CL, where $X$ denotes any possible particle combination~\cite{BESIII:2020iwk}. Considering that these and other searches are almost background-free, one expects significant improvement in sensitivity at next-generation $J/\psi$ factories. Rare hyperon decay measurements are also carried out by the LHCb collaboration. The first evidence for the for the $\Sigma^+\to p\mu^+\mu^-$ decay has been reported with the Run~1 dataset, leading to ${\cal B}(\Sigma^+\to p\mu^+\mu^-) = \left(2.2_{-1.3}^{+1.8}\right)\times 10^{-8}$~\cite{LHCb:2017rdd} consistent with the SM expectation and inconsistent with the HyperCP anomaly~\cite{HyperCP:2005mvo}. Future LHCb datasets promise a broad hyperon decay program, including measurements of semileptonic decays, as well as lepton universality, baryon and lepton number conservation tests~\cite{AlvesJunior:2018ldo}.

In summary, hyperons provide complementary information to kaons in terms of probing BSM physics. The two current experiments, BESIII and LHCb, are exploiting different methodologies for CP asymmetry and rare decay measurements, and offer future prospects.

%% file: rf2_pions.tex
\section{Charged pion decays}
\label{sec:rf2:pions}


Relatively simple dynamics, the small number of available decay channels, and the well controlled radiative and loop corrections make charged pion decays a sensitive means for testing the underlying symmetries and the universality of weak fermion couplings, as well as for improving the understanding of chiral dynamics. Charged pion decays have provided important early insight into the $V-A$ nature of the weak interaction following the discovery of the suppression of the $\pi^+\to e^+\nu$ decay. The ratio of leptonic decay rates of the charged pion, $R_{e\mu}={\cal B}(\pi^+\to e^+\nu)/{\cal B}(\pi^+\to\mu^+\nu)$, is computed within the SM to an exceptional accuracy due to the cancellation of hadronic effects, $R_{e\mu}^{\rm SM}=1.23524(15)\times 10^{-4}$~\cite{Bryman:2021teu}. Precision measurements of this quantity are of high interest, leading to sensitive lepton universality tests. The current world average of $R_{e\mu}^{\rm exp}=1.2327(23)\times 10^{-4}$~\cite{ParticleDataGroup:2020ssz}, dominated by a measurement from the PIENU experiment at TRIUMF~\cite{PiENu:2015seu}, has a remarkable 0.2\% precision.

Phase I of the PIONEER experiment approved at the Paul Scherrer Institute (PSI)~\cite{PIONEER:2022yag,PIONEER:2022alm}, motivated in part by the recent hints for lepton flavor universality violation~\cite{Fischer:2021sqw},
aims to measure the quantity $R_{e\mu}$ to a $10^{-4}$ precision based on a sample of $2\times 10^8$ $\pi^+\to e^+\nu$ decays to be collected in three years of operation, thereby improving the current world average by a factor of about~15~\cite{ParticleDataGroup:2020ssz}. This would match the accuracy of the SM calculation~\cite{Bryman:2021teu}, thereby probing effects of new particles up to the PeV mass scale~\cite{Bryman:2011zz}. The schedule includes extensive detector R\&D efforts, and full-scale physics operation can be expected to start in 2029.

The design of the PIONEER experiment is based on the experience from its predecessors PIENU at TRIUMF~\cite{PiENu:2015seu} and PEN/PIBETA at PSI~\cite{Pocanic:2014jka,PEN:2018kgj}, and is optimized for handling of high beam rates and for reduction of the systematic uncertainties. The layout involves a high-intensity $\pi^+$ beam brought to rest in a highly segmented active target, a LXe electromagnetic calorimeter surrounding the target and providing a $3\pi$~sr solid-angle coverage, as well as beam and positron trackers.

Phase I of the experiment will operate with a continuous high-intensity (300~kHz), low-momentum (55~MeV/$c$) pion beam focused to a small spot size within the target dimensions. The low beam momentum allows for a separator to be used to reduce the contamination of muons and positrons in the beam to a level below 10\%. The segmented active target based on Low Gain Avalanche Detector (LGAD) technology~\cite{Mazza:2021adt} represents a key new feature of the experiment. The target design involves 50~layers (of $2\times 2$~cm$^2$ dimensions) of silicon strip sensors positioned transverse to the beam, at two perpendicular strip orientations, with a 200~$\mu$m pitch and a 120~$\mu$m thickness along the beam. The target will provide excellent resolutions on both position and time (100~ps), thus allowing for reduction of backgrounds from beam pion and muon decays in flight based on the energy deposit pattern, and minimization of acceptance-related systematic effects.

The monoenergetic positron from $\pi^+\to e^+\nu$ decays of the stopped pions has an energy of 69.3~MeV. On the other hand, positrons from the decay chain $\pi^+\to\mu^+\nu$, $\mu^+\to e^+\nu\bar\nu$ form the Michel spectrum with the endpoint of 52.3~MeV. Therefore the majority of positrons from $\pi^+\to e^+\nu$ decays are well isolated from the Michel spectrum and can be identified using a high-resolution calorimeter. The principal challenge lies in determining the low-energy tail of the electromagnetic shower, located under the Michel spectrum. A liquid xenon (LXe) calorimeter read out by UV-sensitive phototubes and vacuum ultraviolet silicon photomultipliers has been proposed, drawing on the experience from the MEG and \mbox{MEG-II} experiments~\cite{Mihara:2011zza,MEG:2016leq,MEGII:2018kmf}. This option provides fast timing, high light yield, uniform response and excellent energy resolution (1.5\% at 70~MeV). A LXe calorimeter of $25X_0$ thickness is expected to provide an intrinsic positron energy tail fraction below 52~MeV at the 0.5\% level for signal events (to be compared to 3\% achieved by PIENU with a $19X_0$ thick calorimeter). The tail fraction can be measured to a $10^{-2}$ precision in relative terms, using control datasets. Therefore the systematic uncertainty in the $R_{e\mu}$ measurement due to the energy tail correction, which amounted to 0.12\% and represented a fundamental limitation of the PIENU experiment~\cite{PiENu:2015seu}, will be reduced to a level below $10^{-4}$.

The broader physics program of PIONEER Phase-I includes searches for ultra-rare $\pi^+$ decay modes, searches for heavy neutral lepton production in $\pi^+\to\ell^+N$ decays improving by an order-of-magnitude on the PIENU experiment~\cite{PIENU:2017wbj,Bryman:2019bjg,PIENU:2019usb} in terms of sensitivity to the coupling parameters $|U_{\ell 4}|^2$ as illustrated for the electron coupling in Fig.~\ref{fig:rf2:hnl}, a search for $\pi^+\to e^+\nu X$ decay involving axions or hidden sectors, and a search for production of a feebly interacting neutral boson produced in $\mu^+\to e^+X$ decay.

In the longer term, Phase II of the PIONEER experiment will focus on the measurement of pion beta decay ($\pi^+\to\pi^0e^+\nu$) rate to 0.2\% precision, improving by a factor of three over the measurement performed by the dedicated PIBETA experiment, ${\cal B}=1.036(6)\times 10^{-8}$~\cite{Pocanic:2003pf}, which currently dominates the world average. This process offers the theoretically cleanest determination of the CKM element $V_{ud}$: thanks to the recent progress in lattice QCD calculations of the radiative corrections~\cite{Feng:2020zdc}, the intrinsic theory uncertainty in the $V_{ud}$ extraction is now reduced to the 0.01\% level. The excellent theory precision allows for a CKM unitarity test~\cite{Bryman:2021teu,Czarnecki:2019iwz}, which is of high interest in the context of the Cabibbo angle anomaly (Fig.~\ref{fig:VusVud}).
The challenge of the $\pi^+\to\pi^0e^+\nu$ measurement is determined by the low branching ratio of ${\cal O}(10^{-8})$. To collect a sample of $7\times 10^{5}$ signal events in 4--5 years of operation during Phase II, an increase in the beam rate to above 10~MHz is required, which can be achieved with larger beam momentum and emittance than in Phase I. The $\pi^+\to\pi^0e^+\nu$ decay will be detected by observing the characteristic nearly back-to-back photons from $\pi^0\to\gamma\gamma$ decay, normalizing to the $\pi^+\to e^+\nu$ decay.

Phase III of the experiment aims for another factor of three improvement in the $\pi^+\to\pi^0 e^+\nu$ rate measurement. The ultimate precision on $|V_{ud}|$ is expected to reach the 0.02\% level, which is comparable to the currently most precise determination from superallowed beta decays. Collection of $7\times 10^6$ signal events would be required for this measurement.

In summary, an ambitious $\pi^+$ decay research program spanning more than a decade has been approved at PSI,
addressing the key issues in the quark flavor sector (lepton universality and CKM unitarity)  at a new level of precision. The US groups are leading proponents of the project, with a strong expertise in the key detector systems. Support of the US effort at both universities and national labs is crucial.

%% file: rf2_eta.tex
\boldmath
\section{Decays of
$\eta^{(\prime)}$ mesons}
\unboldmath

\label{sec:rf2:eta}

\subsection{Theory overview}
\label{sec:rf2:eta:theory}

The $\eta^{(\prime)}$ mesons have the quantum numbers of the vacuum and the Higgs boson, except for parity. This is a rare occurrence in nature, which not only constrains the decay dynamics of these mesons, but also provides a pure initial state free of any SM charge. Strong and electromagnetic $\eta^{(\prime)}$ decays are either anomalous or forbidden at the lowest order due to symmetries and angular momentum conservation, which enhances the relative importance of higher-order contributions and new weakly-coupled interactions. Therefore $\eta^{(\prime)}$ decays provide a unique, flavor-conserving laboratory for investigation of fundamental physics under a broad theme of symmetry and symmetry-breaking, both within the SM and beyond. The highest-priority decay channels~\cite{Gan:2020aco}, allowing for precision SM tests, searches for BSM physics, discrete symmetry violation or potential new light particles, are listed in Table~\ref{tab:summary}.
It should be noted that, with a sufficiently large sample of $\eta^{(\prime)}$ meson decays, all the four portals connecting the SM with possible dark sectors could be explored, including the vector~\cite{Feng:2016jff,Nelson:1989fx,Tulin:2014tya}, (pseudo)scalar~\cite{Batell:2018fqo,Liu:2018qgl} and heavy neutral lepton portals.


\vspace{2mm}
{\bf Standard Model decays}
\begin{itemize}
\item The $\eta\to 3\pi$ decay provides the only experimental determination of the light quark mass difference, $m_d - m_u$. This decay is forbidden by isospin symmetry: three pions cannot combine into a system with vanishing angular momentum, zero isospin, and even C-parity. Isospin-breaking contributions can arise in the SM from either electromagnetic or strong interactions. The \mbox{$\eta\to 3\pi$} decay is unique because
the electromagnetic contributions vanish at leading order in Chiral Perturbation Theory ($\chi$PT)~\cite{Bell:1968wta,Sutherland:1966zz}, and higher order contributions are also suppressed~\cite{Baur:1995gc,Ditsche:2008cq}, providing direct access to $m_d-m_u$ via the quark mass double ratio
\mbox{$Q =(m_s^2-\hat{m}^2)/(m_d^2-m_u^2)$}, with $\hat{m} = (m_u+m_d)/2$. Theoretical issues to be addressed concern a more precise matching to the $\chi$PT representation, which currently constitutes the dominating theoretical uncertainty~\cite{Colangelo:2018jxw}, and an improved implementation of radiative corrections and other higher-order isospin-breaking 
effects~\cite{Ditsche:2008cq,Schneider:2010hs}.
In principle, $\eta^\prime\to3\pi$ can also give access to the light quark mass difference. However the interpretation of $\eta^\prime\to3\pi$ measurements requires significant theoretical advances: while a dispersive representation to describe the Dalitz plot distributions is 
within reach, its matching to a systematic effective field theory comparable to the program for $\eta\to3\pi$ ($\chi$PT, even if supplemented with large-$N_c$ arguments) requires further development.

\item Precision measurements of the $\eta^{(\prime)}\to\gamma\gamma$ decay widths are of importance for several reasons: they serve as reference channels for other decays;
they provide the normalization of the corresponding transition form factors and thus
contribute to our understanding of hadronic light-by-light scattering in \mbox{$(g-2)_\mu$}~\cite{Colangelo:2022jxc}; and their pattern allows the extraction of $\eta$--$\eta^\prime$ mixing parameters, improving the understanding of the $U(1)_A$ anomaly and its interplay with chiral symmetry.

\item The $\eta^{(\prime)}$ transition form factors both have an impact on hadronic contributions to $(g-2)_\mu$, and provide insight into the hadronic structure. Data in different kinematic regimes, from all possible decay and production processes, should be analyzed simultaneously, with theoretical representations
apt for such analytic continuation.  Besides direct measurements, the statistical leverage of using hadronic
decay channels to reconstruct the transition form factors dispersively should be taken advantage of, 
in particular via data on decays such as $\eta^\prime\to2(\pi^+\pi^-)$,
$\eta^\prime\to\pi^+\pi^-e^+e^-$, $\eta^\prime\to\omega e^+e^-$ (or, in a production reaction, $e^+e^-\to\eta\pi^+\pi^-$).
Detailed differential information represents the primary goal.

\item The $\eta\to\pi^0\gamma\gamma$ decay: while decay amplitude suppression in the chiral expansion and the interplay between vector exchange and scalar $S$-wave rescattering dynamics make for a highly interesting theoretical challenge to predict its properties accurately, the interpretation in terms of fundamental insights needs to be improved. This includes a simultaneous analysis of $\eta\to\pi^0\gamma\gamma$, $\eta^\prime\to\eta\gamma\gamma$ and $\eta^\prime\to\pi^0\gamma\gamma$ decays, and the relation to hadronic light-by-light scattering. In addition to high-precision
investigation of QCD dynamics, these processes also allow a search for new light particles.
\end{itemize}


\begin{table}
\centering
\renewcommand{\arraystretch}{1.3}
\begin{tabular}{lccc}
\hline
Decay channel & Standard Model & $\!\!\!$Discrete symmetries$\!\!\!$ & Light BSM particles \\
\hline
$\eta\to\pi^+\pi^-\pi^0$ & Light quark masses & C/CP violation & Scalar bosons \\
$\eta^{(\prime)}\to\gamma\gamma$ & $\eta$--$\eta^\prime$ mixing, partial widths & & \\
$\eta^{(\prime)}\to\ell^+\ell^-\gamma$  & $(g-2)_\mu$ & LFU & $Z'$ bosons, dark photon \\
$\eta\to\pi^0\gamma\gamma$ & $\!\!$Higher-order $\chi$PT, scalar dynamics$\!\!$ & &  $U(1)_B$ boson, scalar bosons\\
$\eta^{(\prime)}\to\mu^+\mu^-$ & $(g-2)_\mu$, precision tests & CP violation & \\
$\eta\to\pi^0\ell^+\ell^-$ & & C violation & Scalar bosons \\
$\eta^{(\prime)}\to\pi^+\pi^- \ell^+\ell^-\!\!$ & $(g-2)_\mu$ &  LFU & ALPs, dark photon \\
$\eta^{(\prime)}\to\pi^0\pi^0 \ell^+\ell^-$ & & C violation, LFU & ALPs \\
\hline
\end{tabular}
\renewcommand{\arraystretch}{1.0}
\caption{High-priority $\eta^{(\prime)}$ decays, with emphasis on synergies of SM and BSM investigations~\cite{Gan:2020aco}.
\label{tab:summary}}
\end{table}


\newpage

{\bf Symmetry tests and lepton flavor/universality violation}
\begin{itemize}
\item P and CP violation. From the theory point of view, a large number of P and CP-violating $\eta^{(\prime)}$ decays are excluded indirectly through the stringent neutron EDM bounds. An exception is the muon polarization asymmetry in the $\eta\to\mu^+\mu^-$ decay probing flavor-conserving CP violation in the second generation~\cite{Sanchez-Puertas:2018tnp}.
Improved experimental bounds on classic channels such as $\eta^{(\prime)}\to\pi\pi$ are welcome but unlikely to find evidence of BSM physics in the foreseeable future.
\item C and CP violation. On the theory side, much of the literature on C and CP violation in $\eta^{(\prime)}$ decays, particularly in regards to the appearance of mirror symmetry breaking in the $\eta\to\pi^+ \pi^-\pi^0$ Dalitz plot,    predates the SM, and little has been done within a modern context. The C and CP-violating charge asymmetry in the Dalitz plot appears through the interference of a new-physics amplitude with the SM, making it linear, rather than quadratic, in C and CP-violating parameters~\cite{Gardner:2019nid}. The manner in which this effect can be generated has been studied in SM effective field theory (SMEFT), and it can be related to operators not participating in the generation of a permanent EDM~\cite{Shi:2020ffh,REDTOP:2022slw}. In contrast, given EDM constraints, the traditional C-violating channel $\eta^{(\prime)} \to 3\gamma$ probes BSM physics that is C and P-violating but CP-conserving. It should be investigated further how large the $\eta^{(\prime)} \to 3\gamma$ rate could be in such a scenario. On the experimental side, the study of C and CP-violating asymmetries in the $\eta\to\pi^+\pi^-\pi^0$ Dalitz plot offers synergies with the QCD/SM motivation for high-precision investigations of this channel~\cite{Gardner:2019nid,Akdag:2021efj}. Channels originally proposed to test C and CP-violation can also be used to search for new light particles. These include $\eta^{(\prime)}\to\pi^0(\pi^0)\ell^+\ell^-$ which are C and CP-violating as single-$\gamma$ processes in QED, but can arise from new light (pseudo)scalar bosons.
\item Second class currents in $\tau$ decays, such as $\tau^-\to\pi^-\eta^{(\prime)}\nu$, are suppressed in the SM by isospin-violating effects to a few parts in $10^{-5}$. This offers an interesting case to search for BSM physics, as the rates of $\tau^-\to\pi^-\eta^{(\prime)}\nu$ decays are enhanced by the contribution of scalar currents from a putative extended Higgs sector~\cite{Jung:2010ik} or leptoquark bosons~\cite{Becirevic:2016yqi}.  
A precision measurement of the $\tau^-\to\pi^-\eta^{(\prime)}\nu$ branching fraction expected at Belle~II~\cite{Belle-II:2022cgf} would put stringent constraints on non-SM scalar interactions, stronger than those from other low-energy observables~\cite{Garces:2017jpz}, provided theoretical progress in the determination of the scalar form factor is made.
\item Lepton flavor violation.
In light of strong constraints on $\mu \to e$ conversion on nuclei, further theoretical studies are required to motivate the searches for the lepton flavor violating $\eta^{(\prime)} \to e^\pm \mu^\mp$ decays. On the other hand, decays that violate charged lepton flavor by two units, such as $\eta^{(\prime)} \to e^\pm e^\pm \mu^\mp \mu^\mp$, are not similarly constrained.
\item
Lepton flavor universality (LFU). Leptonic and semileptonic decays of the $\eta^{(\prime)}$ mesons have clear experimental signatures, and represent a good opportunity for LFU tests. The most relevant groups of processes are $\eta^{(\prime)}\to\ell_1\bar\ell_1\ell_2\bar\ell_2$ and $\eta^{(\prime)}\to\gamma\ell\bar\ell$. The decay amplitude is dominated by the $\eta^{(\prime)}\to\gamma\gamma$ vertex in both cases.
This contribution may be affected by BSM physics, for example $t$-channel exchange of a leptoquark.
%
\end{itemize}

\newpage

\boldmath
\subsection{Planned $\eta^{(\prime)}$ decay experiments}
\unboldmath
\label{sec:rf2:eta:exp}

The JEF experiment~\cite{pro-JEF14,pro-JEF17} under construction at Jefferson Laboratory aims for precision measurements of a wide range of $\eta^{(\prime)}$ decays, with emphasis on rare decay modes. The experiment was approved in 2017, and the data collection is expected to start in 2024. The JEF physics program includes improved measurements of SM $\eta^{(\prime)}$ decays and tests of discrete symmetries, as discussed in Section~\ref{sec:rf2:eta:theory}. 
Highly boosted $\eta^{(\prime)}$ mesons will be produced at the JEF experiment by a 8--12~GeV tagged photon beam via the $\gamma +p\to\eta^{(\prime)} +p$ reaction, at a rate of $5\times 10^7$ of both $\eta$ and $\eta^\prime$ species in 100 days of operation. The experiment will use an upgraded GlueX setup~\cite{GlueX:2020idb}. In particular, the new forward electromagnetic calorimeter (FCAL-II) will involve a new high-granularity array of 1600 PbWO$_4$ scintillating crystals of $2\times 2$~cm$^{2}$ cross-section in the central region. A prototype array of 140~crystals was tested successfully at Jefferson Laboratory in 2019~\cite{Asaturyan:2021ese}, and FCAL-II construction is currently in progress. Photons and leptons from $\eta^{(\prime)}$ decays will be measured by the FCAL-II calorimeter, while charged $\eta^{(\prime)}$ decay products will be measured by the GlueX solenoid magnetic spectrometer and time-of-flight detectors. Backgrounds will be suppressed thanks to $\eta^{(\prime)}$ tagging via recoil proton detection. The highly-boosted $\eta^{(\prime)}$ mesons and the tagging capability offers two orders of magnitude improvement in background suppression and improved control of systematic effects with respect to earlier experiments~\cite{Adlarson:2019nwa, Kashevarov:2017kqb, Husken:2019dou, Berlowski:2019hst, Shuang-shiFang:2015vva}.

In the longer term, the proposed  next-generation REDTOP $\eta^{(\prime)}$ factory aims to produce $\mathcal{O}(10^{14})$ $\eta$ mesons and $\mathcal{O}(10^{12})$ $\eta^{\prime}$ mesons in a
few years of operation~\cite{REDTOP:2022slw}, addressing at a new level of precision rare $\eta^{(\prime)}$ decays, BSM physics, and searches for the violation of fundamental symmetries or lepton flavor universality. From the experimental point of view, REDTOP is complementary to JEF: it is designed to final states with at least two charged particles, as opposed to neutral particles. Recent studies indicate that the experiment could reach sensitivity to the branching ratios of certain BSM processes of the order $10^{-11}$. The REDTOP proposal involves production of $\eta^{(\prime)}$ mesons via decays of baryonic resonances by nuclear scattering of a continuous-wave proton beam in a Lithium target composed of multiple thin foils. Proton kinetic energy of 1.8~(3.6)~GeV is proposed for $\eta$ ($\eta^\prime$) production. In the $\eta$ mode, the required proton rate of $10^{11}$~POT/second (corresponding to 30~W beam power) would generate an inelastic interaction rate of 1~GHz and produce $5.1\times 10^6$ $\eta$ mesons per second. The proposed REDTOP setup employs a 4$\pi$ geometry surrounding the target, and includes a vertex detector capable of identifying secondary vertices, a central tracker, a threshold Cherenkov detector used for particle identification and time-of-flight measurements, and a dual-readout sampling electromagnetic calorimeter. The $\eta^{(\prime)}$ mesons will be produced almost at rest, therefore minimization of the material budget in the detector is a primary consideration (in particular, the expected total beam pipe and tracker material budget is below 1\%$X_0$). The high inelastic interaction rate poses significant challenges for timing, trigger and data processing. Although the choice of technologies for the trackers and the facility hosting the experiment are still to be finalized (FNAL, CERN, BNL, GSI, ESS and HIAF have been considered), the experiment aims to start taking data within the next decade. 


In summary, $\eta$ and $\eta^{\prime}$ mesons provide a unique, flavor-conserving laboratory for tests of low-energy quantum chromodynamics and searches for BSM physics, and there is a strong physics case for $\eta$ factories. The JEF experiment is expected to improve significantly the precision on $\eta^{(\prime)}$ decay measurements in the next decade. Design and sensitivity studies are in progress for a next-generation $\eta^{(\prime)}$ factory, REDTOP.

%% file: rf2_summary.tex
\section{Summary}
\label{sec:rf2:summary}

Precision measurements of strange and light quark decays represent an essential element in the international particle physics program.
Present and next-generation experiments, performed at high-intensity facilities, have a unique potential to observe signatures of new physics beyond the SM description at mass scales well beyond those directly accessible by current or foreseeable accelerators (up to the PeV scale), or to discover hidden sectors (with mediator masses below 1~GeV). Major progress in kaon, hyperon, pion, and $\eta^{(\prime)}$ decay measurements is foreseen in the next decade, and progress in theory and lattice QCD will be crucial for the interpretation of the experimental results. 

It is recommended that the US particle physics community maintains and extends the involvement with this vital area of research, taking advantage of the medium-scale initiatives, many of which are centered in Europe and Asia. These experiments lead to powerful physics insights on relatively short time scales, offer opportunities of making leading contributions to potential BSM discoveries, and provide comprehensive experimental training at various stages including experimental design, R\&D, detector construction, data collection and analysis.

%% file: _main.bbl
\providecommand{\href}[2]{#2}\begingroup\raggedright\begin{thebibliography}{100}

\bibitem{Aebischer:2022vky}
J.~Aebischer, A.J.~Buras and J.~Kumar, \emph{{On the Importance of Rare Kaon
  Decays: A Snowmass 2021 White Paper}},  in \emph{{2022 Snowmass Summer
  Study}}, 3, 2022 [\href{https://arxiv.org/abs/2203.09524}{{\ttfamily
  2203.09524}}].

\bibitem{Buras:2014zga}
A.J.~Buras, D.~Buttazzo, J.~Girrbach-Noe and R.~Knegjens, \emph{{Can we reach
  the Zeptouniverse with rare $K$ and $B_{s,d}$ decays?}},
  \href{https://doi.org/10.1007/JHEP11(2014)121}{\emph{JHEP} {\bfseries 11}
  (2014) 121} [\href{https://arxiv.org/abs/1408.0728}{{\ttfamily 1408.0728}}].

\bibitem{DAmbrosio:2017klp}
G.~D'Ambrosio and T.~Kitahara, \emph{{Direct $CP$ Violation in $K \to \mu^+
  \mu^-$}}, \href{https://doi.org/10.1103/PhysRevLett.119.201802}{\emph{Phys.
  Rev. Lett.} {\bfseries 119} (2017) 201802}
  [\href{https://arxiv.org/abs/1707.06999}{{\ttfamily 1707.06999}}].

\bibitem{Dery:2021mct}
A.~Dery, M.~Ghosh, Y.~Grossman and S.~Schacht, \emph{{K \textrightarrow{}
  \ensuremath{\mu}$^{+}$\ensuremath{\mu}$^{-}$ as a clean probe of
  short-distance physics}},
  \href{https://doi.org/10.1007/JHEP07(2021)103}{\emph{JHEP} {\bfseries 07}
  (2021) 103} [\href{https://arxiv.org/abs/2104.06427}{{\ttfamily
  2104.06427}}].

\bibitem{Cirigliano:2011ny}
V.~Cirigliano, G.~Ecker, H.~Neufeld, A.~Pich and J.~Portoles, \emph{{Kaon
  Decays in the Standard Model}},
  \href{https://doi.org/10.1103/RevModPhys.84.399}{\emph{Rev. Mod. Phys.}
  {\bfseries 84} (2012) 399} [\href{https://arxiv.org/abs/1107.6001}{{\ttfamily
  1107.6001}}].

\bibitem{Buras:2015qea}
A.J.~Buras, D.~Buttazzo, J.~Girrbach-Noe and R.~Knegjens, \emph{{$ {K}^{+}\to
  {\pi}^{+}\nu \overline{\nu} $ and $ {K}_L\to {\pi}^0\nu \overline{\nu} $ in
  the Standard Model: status and perspectives}},
  \href{https://doi.org/10.1007/JHEP11(2015)033}{\emph{JHEP} {\bfseries 11}
  (2015) 033} [\href{https://arxiv.org/abs/1503.02693}{{\ttfamily
  1503.02693}}].

\bibitem{Buras:2021nns}
A.J.~Buras and E.~Venturini, \emph{{Searching for New Physics in Rare $K$ and
  $B$ Decays without $|V_{cb}|$ and $|V_{ub}|$ Uncertainties}},
  \href{https://arxiv.org/abs/2109.11032}{{\ttfamily 2109.11032}}.

\bibitem{Buras:2022wpw}
A.J.~Buras and E.~Venturini, \emph{{The Exclusive Vision of Rare $K$ and $B$
  Decays and of the Quark Mixing in the Standard Model}},
  \href{https://arxiv.org/abs/2203.11960}{{\ttfamily 2203.11960}}.

\bibitem{Buras:2022qip}
A.J.~Buras, \emph{{Standard Model Predictions for Rare K and B Decays without
  New Physics Infection}},  \href{https://arxiv.org/abs/2209.03968}{{\ttfamily
  2209.03968}}.

\bibitem{Bai:2018hqu}
Z.~Bai, N.H.~Christ, X.~Feng, A.~Lawson, A.~Portelli and C.T.~Sachrajda,
  \emph{{$K^+\to\pi^+\nu\bar{\nu}$ decay amplitude from lattice QCD}},
  \href{https://doi.org/10.1103/PhysRevD.98.074509}{\emph{Phys. Rev. D}
  {\bfseries 98} (2018) 074509}
  [\href{https://arxiv.org/abs/1806.11520}{{\ttfamily 1806.11520}}].

\bibitem{Christ:2019dxu}
{\scshape RBC, UKQCD} collaboration, \emph{{Lattice QCD study of the rare kaon
  decay $K^+\to\pi^+\nu\bar{\nu}$ at a near-physical pion mass}},
  \href{https://doi.org/10.1103/PhysRevD.100.114506}{\emph{Phys. Rev. D}
  {\bfseries 100} (2019) 114506}
  [\href{https://arxiv.org/abs/1910.10644}{{\ttfamily 1910.10644}}].

\bibitem{Blum:2022wsz}
T.~Blum et~al., \emph{{Discovering new physics in rare kaon decays}},  in
  \emph{{2022 Snowmass Summer Study}}, 3, 2022
  [\href{https://arxiv.org/abs/2203.10998}{{\ttfamily 2203.10998}}].

\bibitem{Buras:2015yca}
A.J.~Buras, D.~Buttazzo and R.~Knegjens, \emph{{$ K\to \pi \nu \overline{\nu} $
  and \ensuremath{\varepsilon}'/\ensuremath{\varepsilon} in simplified new
  physics models}}, \href{https://doi.org/10.1007/JHEP11(2015)166}{\emph{JHEP}
  {\bfseries 11} (2015) 166}
  [\href{https://arxiv.org/abs/1507.08672}{{\ttfamily 1507.08672}}].

\bibitem{Buchalla:1994tr}
G.~Buchalla and A.J.~Buras, \emph{{$\sin2\beta$ from $K\to\pi\nu\bar\nu$}},
  \href{https://doi.org/10.1016/0370-2693(94)91034-0}{\emph{Phys. Lett. B}
  {\bfseries 333} (1994) 221}
  [\href{https://arxiv.org/abs/hep-ph/9405259}{{\ttfamily hep-ph/9405259}}].

\bibitem{Buras:2015jaq}
A.J.~Buras, \emph{{New physics patterns in $\varepsilon^\prime/\varepsilon$ and
  $\varepsilon_K$ with implications for rare kaon decays and $\Delta M_K$}},
  \href{https://doi.org/10.1007/JHEP04(2016)071}{\emph{JHEP} {\bfseries 04}
  (2016) 071} [\href{https://arxiv.org/abs/1601.00005}{{\ttfamily
  1601.00005}}].

\bibitem{Aebischer:2020mkv}
J.~Aebischer, A.J.~Buras and J.~Kumar, \emph{{Another SMEFT story: $Z'$ facing
  new results on $\epsilon'/\epsilon$, $\Delta M_{K}$ and $K \to \pi \nu
  \overline{\nu} $}},
  \href{https://doi.org/10.1007/JHEP12(2020)097}{\emph{JHEP} {\bfseries 12}
  (2020) 097} [\href{https://arxiv.org/abs/2006.01138}{{\ttfamily
  2006.01138}}].

\bibitem{Altmannshofer:2009ma}
W.~Altmannshofer, A.J.~Buras, D.M.~Straub and M.~Wick, \emph{{New strategies
  for New Physics search in $B \to K^{*} \nu \bar{\nu}$, $B \to K \nu
  \bar{\nu}$ and $B \to X_{s} \nu \bar{\nu}$ decays}},
  \href{https://doi.org/10.1088/1126-6708/2009/04/022}{\emph{JHEP} {\bfseries
  04} (2009) 022} [\href{https://arxiv.org/abs/0902.0160}{{\ttfamily
  0902.0160}}].

\bibitem{Buras:2014fpa}
A.J.~Buras, J.~Girrbach-Noe, C.~Niehoff and D.M.~Straub, \emph{{$ B\to
  {K}^{\left(\ast \right)}\nu \overline{\nu} $ decays in the Standard Model and
  beyond}}, \href{https://doi.org/10.1007/JHEP02(2015)184}{\emph{JHEP}
  {\bfseries 02} (2015) 184} [\href{https://arxiv.org/abs/1409.4557}{{\ttfamily
  1409.4557}}].

\bibitem{Bordone:2017lsy}
M.~Bordone, D.~Buttazzo, G.~Isidori and J.~Monnard, \emph{{Probing Lepton
  Flavour Universality with $K \to \pi \nu \bar\nu$ decays}},
  \href{https://doi.org/10.1140/epjc/s10052-017-5202-1}{\emph{Eur. Phys. J. C}
  {\bfseries 77} (2017) 618}
  [\href{https://arxiv.org/abs/1705.10729}{{\ttfamily 1705.10729}}].

\bibitem{Deppisch:2020oyx}
F.F.~Deppisch, K.~Fridell and J.~Harz, \emph{{Constraining lepton number
  violating interactions in rare kaon decays}},
  \href{https://doi.org/10.1007/JHEP12(2020)186}{\emph{JHEP} {\bfseries 12}
  (2020) 186} [\href{https://arxiv.org/abs/2009.04494}{{\ttfamily
  2009.04494}}].

\bibitem{Christ:2015phf}
N.H.~Christ and Z.~Bai, \emph{{Computing the long-distance contributions to
  $\varepsilon_K$}}, \href{https://doi.org/10.22323/1.251.0342}{\emph{PoS}
  {\bfseries LATTICE2015} (2016) 342}.

\bibitem{RBC:2020kdj}
{\scshape RBC, UKQCD} collaboration, \emph{{Direct CP violation and the $\Delta
  I=1/2$ rule in $K\to\pi\pi$ decay from the standard model}},
  \href{https://doi.org/10.1103/PhysRevD.102.054509}{\emph{Phys. Rev. D}
  {\bfseries 102} (2020) 054509}
  [\href{https://arxiv.org/abs/2004.09440}{{\ttfamily 2004.09440}}].

\bibitem{Wang:2021twm}
B.~Wang, \emph{{Lattice calculation of the mass difference between the long-
  and short-lived K mesons for physical quark masses}}, Ph.D. thesis, Columbia
  U. (main), 2021.
\newblock 10.7916/d8-fmzr-ys91.

\bibitem{NA48:2002tmj}
{\scshape NA48} collaboration, \emph{{A Precision measurement of direct CP
  violation in the decay of neutral kaons into two pions}},
  \href{https://doi.org/10.1016/S0370-2693(02)02476-0}{\emph{Phys. Lett. B}
  {\bfseries 544} (2002) 97}
  [\href{https://arxiv.org/abs/hep-ex/0208009}{{\ttfamily hep-ex/0208009}}].

\bibitem{KTeV:2010sng}
{\scshape KTeV} collaboration, \emph{{Precise Measurements of Direct CP
  Violation, CPT Symmetry, and Other Parameters in the Neutral Kaon System}},
  \href{https://doi.org/10.1103/PhysRevD.83.092001}{\emph{Phys. Rev. D}
  {\bfseries 83} (2011) 092001}
  [\href{https://arxiv.org/abs/1011.0127}{{\ttfamily 1011.0127}}].

\bibitem{Aoki:2021kgd}
Y.~Aoki et~al., \emph{{FLAG Review 2021}},
  \href{https://arxiv.org/abs/2111.09849}{{\ttfamily 2111.09849}}.

\bibitem{Hardy:2020qwl}
J.C.~Hardy and I.S.~Towner, \emph{{Superallowed $0^+ \to 0^+$ nuclear $\beta$
  decays: 2020 critical survey, with implications for V$_{ud}$ and CKM
  unitarity}}, \href{https://doi.org/10.1103/PhysRevC.102.045501}{\emph{Phys.
  Rev. C} {\bfseries 102} (2020) 045501}.

\bibitem{Czarnecki:2018okw}
A.~Czarnecki, W.J.~Marciano and A.~Sirlin, \emph{{Neutron Lifetime and Axial
  Coupling Connection}},
  \href{https://doi.org/10.1103/PhysRevLett.120.202002}{\emph{Phys. Rev. Lett.}
  {\bfseries 120} (2018) 202002}
  [\href{https://arxiv.org/abs/1802.01804}{{\ttfamily 1802.01804}}].

\bibitem{Seng:2020wjq}
C.-Y.~Seng, X.~Feng, M.~Gorchtein and L.-C.~Jin, \emph{{Joint lattice
  QCD\textendash{}dispersion theory analysis confirms the quark-mixing top-row
  unitarity deficit}},
  \href{https://doi.org/10.1103/PhysRevD.101.111301}{\emph{Phys. Rev. D}
  {\bfseries 101} (2020) 111301}
  [\href{https://arxiv.org/abs/2003.11264}{{\ttfamily 2003.11264}}].

\bibitem{Cirigliano:2022yyo}
V.~Cirigliano, A.~Crivellin, M.~Hoferichter and M.~Moulson, \emph{{Scrutinizing
  CKM unitarity with a new measurement of the $K_{\mu 3}/K_{\mu 2}$ branching
  fraction}},  \href{https://arxiv.org/abs/2208.11707}{{\ttfamily 2208.11707}}.

\bibitem{Bryman:2021teu}
D.~Bryman, V.~Cirigliano, A.~Crivellin and G.~Inguglia, \emph{{Testing Lepton
  Flavor Universality with Pion, Kaon, Tau, and Beta Decays}},
  \href{https://arxiv.org/abs/2111.05338}{{\ttfamily 2111.05338}}.

\bibitem{Ishikawa:2022otj}
{\scshape PACS} collaboration, \emph{{$K_{\ell 3}$ form factors at the physical
  point towards the continuum limit}},
  \href{https://arxiv.org/abs/2206.08654}{{\ttfamily 2206.08654}}.

\bibitem{Carrasco:2015xwa}
N.~Carrasco, V.~Lubicz, G.~Martinelli, C.T.~Sachrajda, N.~Tantalo, C.~Tarantino
  et~al., \emph{{QED Corrections to Hadronic Processes in Lattice QCD}},
  \href{https://doi.org/10.1103/PhysRevD.91.074506}{\emph{Phys. Rev. D}
  {\bfseries 91} (2015) 074506}
  [\href{https://arxiv.org/abs/1502.00257}{{\ttfamily 1502.00257}}].

\bibitem{Boyle:2019rdx}
P.A.~Boyle, V.~Guelpers, A.~Juettner, C.~Lehner, F.O.~hOgain, A.~Portelli
  et~al., \emph{{QED corrections to leptonic decay rates}},
  \href{https://doi.org/10.22323/1.334.0267}{\emph{PoS} {\bfseries LATTICE2018}
  (2019) 267} [\href{https://arxiv.org/abs/1902.00295}{{\ttfamily
  1902.00295}}].

\bibitem{Feng:2021zek}
X.~Feng, L.~Jin and M.J.~Riberdy, \emph{{Lattice QCD Calculation of the Pion
  Mass Splitting}},
  \href{https://doi.org/10.1103/PhysRevLett.128.052003}{\emph{Phys. Rev. Lett.}
  {\bfseries 128} (2022) 052003}
  [\href{https://arxiv.org/abs/2108.05311}{{\ttfamily 2108.05311}}].

\bibitem{Seng:2020jtz}
C.-Y.~Seng, X.~Feng, M.~Gorchtein, L.-C.~Jin and U.-G.~Mei\ss{}ner, \emph{{New
  method for calculating electromagnetic effects in semileptonic beta-decays of
  mesons}}, \href{https://doi.org/10.1007/JHEP10(2020)179}{\emph{JHEP}
  {\bfseries 10} (2020) 179}
  [\href{https://arxiv.org/abs/2009.00459}{{\ttfamily 2009.00459}}].

\bibitem{Seng:2019lxf}
C.-Y.~Seng, D.~Galviz and U.-G.~Mei\ss{}ner, \emph{{A New Theory Framework for
  the Electroweak Radiative Corrections in $K_{l3}$ Decays}},
  \href{https://doi.org/10.1007/JHEP02(2020)069}{\emph{JHEP} {\bfseries 02}
  (2020) 069} [\href{https://arxiv.org/abs/1910.13208}{{\ttfamily
  1910.13208}}].

\bibitem{Seng:2021boy}
C.-Y.~Seng, D.~Galviz, M.~Gorchtein and U.G.~Mei\ss{}ner, \emph{{High-precision
  determination of the Ke3 radiative corrections}},
  \href{https://doi.org/10.1016/j.physletb.2021.136522}{\emph{Phys. Lett. B}
  {\bfseries 820} (2021) 136522}
  [\href{https://arxiv.org/abs/2103.00975}{{\ttfamily 2103.00975}}].

\bibitem{Seng:2021wcf}
C.-Y.~Seng, D.~Galviz, M.~Gorchtein and U.-G.~Mei\ss{}ner, \emph{{Improved
  $K_{e3}$ radiative corrections sharpen the K$_{\mu 2}$\textendash{}$K_{\ell
  3}$ discrepancy}}, \href{https://doi.org/10.1007/JHEP11(2021)172}{\emph{JHEP}
  {\bfseries 11} (2021) 172}
  [\href{https://arxiv.org/abs/2103.04843}{{\ttfamily 2103.04843}}].

\bibitem{Seng:2022wcw}
C.-Y.~Seng, D.~Galviz, M.~Gorchtein and U.-G.~Mei\ss{}ner, \emph{{Complete
  theory of radiative corrections to K$_{\ell3}$ decays and the V$_{us}$
  update}}, \href{https://doi.org/10.1007/JHEP07(2022)071}{\emph{JHEP}
  {\bfseries 07} (2022) 071}
  [\href{https://arxiv.org/abs/2203.05217}{{\ttfamily 2203.05217}}].

\bibitem{Belle-II:2022cgf}
{\scshape Belle-II} collaboration, \emph{{Snowmass White Paper: Belle II
  physics reach and plans for the next decade and beyond}},
  \href{https://arxiv.org/abs/2207.06307}{{\ttfamily 2207.06307}}.

\bibitem{Abdullahi:2022jlv}
A.M.~Abdullahi et~al., \emph{{The Present and Future Status of Heavy Neutral
  Leptons}},  in \emph{{2022 Snowmass Summer Study}}, 3, 2022
  [\href{https://arxiv.org/abs/2203.08039}{{\ttfamily 2203.08039}}].

\bibitem{NA62:2020mcv}
{\scshape NA62} collaboration, \emph{{Search for heavy neutral lepton
  production in $K^+$ decays to positrons}},
  \href{https://doi.org/10.1016/j.physletb.2020.135599}{\emph{Phys. Lett. B}
  {\bfseries 807} (2020) 135599}
  [\href{https://arxiv.org/abs/2005.09575}{{\ttfamily 2005.09575}}].

\bibitem{PIENU:2017wbj}
{\scshape PIENU} collaboration, \emph{{Improved search for heavy neutrinos in
  the decay $\pi\rightarrow e\nu$}},
  \href{https://doi.org/10.1103/PhysRevD.97.072012}{\emph{Phys. Rev. D}
  {\bfseries 97} (2018) 072012}
  [\href{https://arxiv.org/abs/1712.03275}{{\ttfamily 1712.03275}}].

\bibitem{Bryman:2019bjg}
D.A.~Bryman and R.~Shrock, \emph{{Constraints on Sterile Neutrinos in the MeV
  to GeV Mass Range}},
  \href{https://doi.org/10.1103/PhysRevD.100.073011}{\emph{Phys. Rev. D}
  {\bfseries 100} (2019) 073011}
  [\href{https://arxiv.org/abs/1909.11198}{{\ttfamily 1909.11198}}].

\bibitem{Crivellin:2016vjc}
A.~Crivellin, G.~D'Ambrosio, M.~Hoferichter and L.C.~Tunstall, \emph{{Violation
  of lepton flavor and lepton flavor universality in rare kaon decays}},
  \href{https://doi.org/10.1103/PhysRevD.93.074038}{\emph{Phys. Rev. D}
  {\bfseries 93} (2016) 074038}
  [\href{https://arxiv.org/abs/1601.00970}{{\ttfamily 1601.00970}}].

\bibitem{Cornella:2019uxs}
C.~Cornella, P.~Paradisi and O.~Sumensari, \emph{{Hunting for ALPs with Lepton
  Flavor Violation}},
  \href{https://doi.org/10.1007/JHEP01(2020)158}{\emph{JHEP} {\bfseries 01}
  (2020) 158} [\href{https://arxiv.org/abs/1911.06279}{{\ttfamily
  1911.06279}}].

\bibitem{Abada:2017jjx}
A.~Abada, V.~De~Romeri, M.~Lucente, A.M.~Teixeira and T.~Toma, \emph{{Effective
  Majorana mass matrix from tau and pseudoscalar meson lepton number violating
  decays}}, \href{https://doi.org/10.1007/JHEP02(2018)169}{\emph{JHEP}
  {\bfseries 02} (2018) 169}
  [\href{https://arxiv.org/abs/1712.03984}{{\ttfamily 1712.03984}}].

\bibitem{Goudzovski:2022vbt}
E.~Goudzovski et~al., \emph{{New Physics Searches at Kaon and Hyperon
  Factories}},  \href{https://arxiv.org/abs/2201.07805}{{\ttfamily
  2201.07805}}.

\bibitem{Krnjaic:2019rsv}
G.~Krnjaic, G.~Marques-Tavares, D.~Redigolo and K.~Tobioka, \emph{{Probing
  Muonphilic Force Carriers and Dark Matter at Kaon Factories}},
  \href{https://doi.org/10.1103/PhysRevLett.124.041802}{\emph{Phys. Rev. Lett.}
  {\bfseries 124} (2020) 041802}
  [\href{https://arxiv.org/abs/1902.07715}{{\ttfamily 1902.07715}}].

\bibitem{KOTO:2022hwn}
{\scshape KOTO, LHCb, NA62/KLEVER, US Kaon Interest Group} collaboration,
  \emph{{Searches for new physics with high-intensity kaon beams}},  4, 2022
  [\href{https://arxiv.org/abs/2204.13394}{{\ttfamily 2204.13394}}].

\bibitem{NA62:2017rwk}
{\scshape NA62} collaboration, \emph{{The Beam and detector of the NA62
  experiment at CERN}},
  \href{https://doi.org/10.1088/1748-0221/12/05/P05025}{\emph{JINST} {\bfseries
  12} (2017) P05025} [\href{https://arxiv.org/abs/1703.08501}{{\ttfamily
  1703.08501}}].

\bibitem{NA62:2021zjw}
{\scshape NA62} collaboration, \emph{{Measurement of the very rare
  K$^{+}$\textrightarrow{}$ {\pi}^{+}\nu \overline{\nu} $ decay}},
  \href{https://doi.org/10.1007/JHEP06(2021)093}{\emph{JHEP} {\bfseries 06}
  (2021) 093} [\href{https://arxiv.org/abs/2103.15389}{{\ttfamily
  2103.15389}}].

\bibitem{BNL-E949:2009dza}
{\scshape BNL-E949} collaboration, \emph{{Study of the decay $K^+\to\pi^+\nu
  \bar\nu$ in the momentum region $140 < P_\pi < 199$ MeV/c}},
  \href{https://doi.org/10.1103/PhysRevD.79.092004}{\emph{Phys. Rev. D}
  {\bfseries 79} (2009) 092004}
  [\href{https://arxiv.org/abs/0903.0030}{{\ttfamily 0903.0030}}].

\bibitem{NA62:2021xxx}
{NA62 Collaboration}, ``2021 {NA62} status report to the {CERN} {SPSC}.'' 2021.

\bibitem{Mescia:2006jd}
F.~Mescia, C.~Smith and S.~Trine, \emph{{$K_L\to\pi^0 e^+e^-$ and
  $K_L\to\pi^0\mu^+\mu^-$: A Binary star on the stage of flavor physics}},
  \href{https://doi.org/10.1088/1126-6708/2006/08/088}{\emph{JHEP} {\bfseries
  08} (2006) 088} [\href{https://arxiv.org/abs/hep-ph/0606081}{{\ttfamily
  hep-ph/0606081}}].

\bibitem{Baldini:2021hfw}
W.~Baldini et~al., \emph{{SHADOWS (Search for Hidden And Dark Objects With the
  SPS)}},  \href{https://arxiv.org/abs/2110.08025}{{\ttfamily 2110.08025}}.

\bibitem{Bartosik:2018xxx}
H.~Bartosik et~al., \emph{{SPS} operation and future proton sharing scenarios
  for the {SHiP} experiment at the {BDF} facility},  Tech. Rep.
  \href{https://cds.cern.ch/record/2650722}{CERN-ACC-NOTE-2018-002}, CERN
  (2018).

\bibitem{Ambrosino:2019qvz}
{\scshape KLEVER Project} collaboration, \emph{{KLEVER: An experiment to
  measure BR($K_L\to\pi^0\nu\bar{\nu}$) at the CERN SPS}},
  \href{https://arxiv.org/abs/1901.03099}{{\ttfamily 1901.03099}}.

\bibitem{Banerjee:2018xxx}
D.~Banerjee et~al., \emph{The report of the {Conventional Beams Working Group}
  to the {Physics Beyond Colliders} study and to the {European Strategy for
  Particle Physics}},  Tech. Rep.
  \href{http://cds.cern.ch/record/2650193/}{CERN-PBC-NOTES-2018-005}, CERN
  (2018).

\bibitem{Gatignon:2018xxx}
L.~Gatignon et~al., \emph{Report from the {Conventional Beams Working Group} to
  the {Physics Beyond Colliders} study and to the {European Strategy for
  Particle Physics}},  Tech. Rep.
  \href{https://cds.cern.ch/record/2650989}{CERN-PBC-REPORT-2018-002}, CERN
  (2018).

\bibitem{VanDijk:2019oml}
M.~Van~Dijk et~al., \emph{{The K12 beamline for the KLEVER experiment}}, .

\bibitem{Beacham:2019nyx}
J.~Beacham et~al., \emph{{Physics Beyond Colliders at CERN: Beyond the Standard
  Model Working Group Report}},
  \href{https://doi.org/10.1088/1361-6471/ab4cd2}{\emph{J. Phys. G} {\bfseries
  47} (2020) 010501} [\href{https://arxiv.org/abs/1901.09966}{{\ttfamily
  1901.09966}}].

\bibitem{KOTO:2018dsc}
{\scshape KOTO} collaboration, \emph{{Search for the $K_L \!\to\! \pi^0 \nu
  \overline{\nu}$ and $K_L \!\to\! \pi^0 X^0$ decays at the J-PARC KOTO
  experiment}},
  \href{https://doi.org/10.1103/PhysRevLett.122.021802}{\emph{Phys. Rev. Lett.}
  {\bfseries 122} (2019) 021802}
  [\href{https://arxiv.org/abs/1810.09655}{{\ttfamily 1810.09655}}].

\bibitem{KOTO:2020prk}
{\scshape KOTO} collaboration, \emph{{Study of the $K_L \to \pi^0 \nu \bar \nu$
  Decay at the J-PARC KOTO Experiment}},
  \href{https://doi.org/10.1103/PhysRevLett.126.121801}{\emph{Phys. Rev. Lett.}
  {\bfseries 126} (2021) 121801}
  [\href{https://arxiv.org/abs/2012.07571}{{\ttfamily 2012.07571}}].

\bibitem{NA62:2019eax}
{\scshape NA62} collaboration, \emph{{Searches for lepton number violating
  $K^+$ decays}},
  \href{https://doi.org/10.1016/j.physletb.2019.07.041}{\emph{Phys. Lett. B}
  {\bfseries 797} (2019) 134794}
  [\href{https://arxiv.org/abs/1905.07770}{{\ttfamily 1905.07770}}].

\bibitem{NA62:2021zxl}
{\scshape NA62} collaboration, \emph{{Search for Lepton Number and Flavor
  Violation in $K^+$ and $\pi^0$ Decays}},
  \href{https://doi.org/10.1103/PhysRevLett.127.131802}{\emph{Phys. Rev. Lett.}
  {\bfseries 127} (2021) 131802}
  [\href{https://arxiv.org/abs/2105.06759}{{\ttfamily 2105.06759}}].

\bibitem{NA62:2022tte}
{\scshape NA62} collaboration, \emph{{Searches for lepton number violating
  $K^+\to\pi^-(\pi^0)e^+e^+$ decays}},
  \href{https://arxiv.org/abs/2202.00331}{{\ttfamily 2202.00331}}.

\bibitem{NA62:2019meo}
{\scshape NA62} collaboration, \emph{{Search for production of an invisible
  dark photon in $\pi^0$ decays}},
  \href{https://doi.org/10.1007/JHEP05(2019)182}{\emph{JHEP} {\bfseries 05}
  (2019) 182} [\href{https://arxiv.org/abs/1903.08767}{{\ttfamily
  1903.08767}}].

\bibitem{NA62:2020xlg}
{\scshape NA62} collaboration, \emph{{Search for a feebly interacting particle
  $X$ in the decay $K^{+}\rightarrow\pi^{+}X$}},
  \href{https://doi.org/10.1007/JHEP03(2021)058}{\emph{JHEP} {\bfseries 03}
  (2021) 058} [\href{https://arxiv.org/abs/2011.11329}{{\ttfamily
  2011.11329}}].

\bibitem{NA62:2020pwi}
{\scshape NA62} collaboration, \emph{{Search for $\pi^0$ decays to invisible
  particles}}, \href{https://doi.org/10.1007/JHEP02(2021)201}{\emph{JHEP}
  {\bfseries 02} (2021) 201}
  [\href{https://arxiv.org/abs/2010.07644}{{\ttfamily 2010.07644}}].

\bibitem{NA62:2021bji}
{\scshape NA62} collaboration, \emph{{Search for $K^+$ decays to a muon and
  invisible particles}},
  \href{https://doi.org/10.1016/j.physletb.2021.136259}{\emph{Phys. Lett. B}
  {\bfseries 816} (2021) 136259}
  [\href{https://arxiv.org/abs/2101.12304}{{\ttfamily 2101.12304}}].

\bibitem{KOTO:2020bhx}
{\scshape KOTO} collaboration, \emph{{First search for $K_L \rightarrow \pi^0
  \gamma$}}, \href{https://doi.org/10.1103/PhysRevD.102.051103}{\emph{Phys.
  Rev. D} {\bfseries 102} (2020) 051103}
  [\href{https://arxiv.org/abs/2006.14918}{{\ttfamily 2006.14918}}].

\bibitem{Fischer:2021sqw}
O.~Fischer et~al., \emph{{Unveiling Hidden Physics at the LHC}},
  \href{https://arxiv.org/abs/2109.06065}{{\ttfamily 2109.06065}}.

\bibitem{ParticleDataGroup:2020ssz}
{\scshape Particle Data Group} collaboration, \emph{{Review of Particle
  Physics}}, \href{https://doi.org/10.1093/ptep/ptaa104}{\emph{PTEP} {\bfseries
  2020} (2020) 083C01}.

\bibitem{NA62:2012lny}
{\scshape NA62} collaboration, \emph{{Precision Measurement of the Ratio of the
  Charged Kaon Leptonic Decay Rates}},
  \href{https://doi.org/10.1016/j.physletb.2013.01.037}{\emph{Phys. Lett. B}
  {\bfseries 719} (2013) 326}
  [\href{https://arxiv.org/abs/1212.4012}{{\ttfamily 1212.4012}}].

\bibitem{Cirigliano:2007xi}
V.~Cirigliano and I.~Rosell, \emph{{Two-loop effective theory analysis of $\pi
  (K)\to e\nu [\gamma]$ branching ratios}},
  \href{https://doi.org/10.1103/PhysRevLett.99.231801}{\emph{Phys. Rev. Lett.}
  {\bfseries 99} (2007) 231801}
  [\href{https://arxiv.org/abs/0707.3439}{{\ttfamily 0707.3439}}].

\bibitem{LHCb:2020ycd}
{\scshape LHCb} collaboration, \emph{{Constraints on the $K^0_S \rightarrow
  \mu^+ \mu^-$ Branching Fraction}},
  \href{https://doi.org/10.1103/PhysRevLett.125.231801}{\emph{Phys. Rev. Lett.}
  {\bfseries 125} (2020) 231801}
  [\href{https://arxiv.org/abs/2001.10354}{{\ttfamily 2001.10354}}].

\bibitem{AlvesJunior:2018ldo}
A.A.~Alves~Junior et~al., \emph{{Prospects for Measurements with Strange
  Hadrons at LHCb}}, \href{https://doi.org/10.1007/JHEP05(2019)048}{\emph{JHEP}
  {\bfseries 05} (2019) 048}
  [\href{https://arxiv.org/abs/1808.03477}{{\ttfamily 1808.03477}}].

\bibitem{HyperCP:2004kbv}
{\scshape HyperCP} collaboration, \emph{{HyperCP: A High-rate spectrometer for
  the study of charged hyperon and kaon decays}},
  \href{https://doi.org/10.1016/j.nima.2004.12.031}{\emph{Nucl. Instrum. Meth.
  A} {\bfseries 541} (2005) 516}
  [\href{https://arxiv.org/abs/hep-ex/0405034}{{\ttfamily hep-ex/0405034}}].

\bibitem{HyperCP:2004zvh}
{\scshape HyperCP} collaboration, \emph{{Search for CP violation in charged-Xi
  and Lambda hyperon decays}},
  \href{https://doi.org/10.1103/PhysRevLett.93.262001}{\emph{Phys. Rev. Lett.}
  {\bfseries 93} (2004) 262001}
  [\href{https://arxiv.org/abs/hep-ex/0412038}{{\ttfamily hep-ex/0412038}}].

\bibitem{BESIII:2022qax}
{\scshape BESIII} collaboration, \emph{{Precision measurements of decay
  parameters and $CP$ asymmetry in $\Lambda$ decays}},
  \href{https://arxiv.org/abs/2204.11058}{{\ttfamily 2204.11058}}.

\bibitem{BESIII:2020fqg}
{\scshape BESIII} collaboration, \emph{{$\Sigma^{+}$ and $\bar{\Sigma}^-$
  polarization in the $J/\psi$ and $\psi(3686)$ decays}},
  \href{https://doi.org/10.1103/PhysRevLett.125.052004}{\emph{Phys. Rev. Lett.}
  {\bfseries 125} (2020) 052004}
  [\href{https://arxiv.org/abs/2004.07701}{{\ttfamily 2004.07701}}].

\bibitem{BESIII:2021ypr}
{\scshape BESIII} collaboration, \emph{{Probing CP symmetry and weak phases
  with entangled double-strange baryons}},
  \href{https://doi.org/10.1038/s41586-022-04624-1}{\emph{Nature} {\bfseries
  606} (2022) 64} [\href{https://arxiv.org/abs/2105.11155}{{\ttfamily
  2105.11155}}].

\bibitem{Salone:2022lpt}
N.~Salone, P.~Adlarson, V.~Batozskaya, A.~Kupsc, S.~Leupold and J.~Tandean,
  \emph{{Study of CP violation in hyperon decays at Super Charm-Tau Factories
  with a polarized electron beam}},  in \emph{{2022 Snowmass Summer Study}}, 3,
  2022 [\href{https://arxiv.org/abs/2203.03035}{{\ttfamily 2203.03035}}].

\bibitem{Tandean:2002vy}
J.~Tandean and G.~Valencia, \emph{{CP violation in hyperon nonleptonic decays
  within the standard model}},
  \href{https://doi.org/10.1103/PhysRevD.67.056001}{\emph{Phys. Rev. D}
  {\bfseries 67} (2003) 056001}
  [\href{https://arxiv.org/abs/hep-ph/0211165}{{\ttfamily hep-ph/0211165}}].

\bibitem{BESIII:2021ynj}
{\scshape BESIII} collaboration, \emph{{First Measurement of the Absolute
  Branching Fraction of $\Lambda \to p \mu^- \bar{\nu}_{\mu}$}},
  \href{https://doi.org/10.1103/PhysRevLett.127.121802}{\emph{Phys. Rev. Lett.}
  {\bfseries 127} (2021) 121802}
  [\href{https://arxiv.org/abs/2107.06704}{{\ttfamily 2107.06704}}].

\bibitem{BESIII:2021emv}
{\scshape BESIII} collaboration, \emph{{Search for the hyperon semileptonic
  decay $\Xi^{-} \rightarrow \Xi^{0} e^{-} \bar{\nu_{e}}$}},
  \href{https://doi.org/10.1103/PhysRevD.104.072007}{\emph{Phys. Rev. D}
  {\bfseries 104} (2021) 072007}
  [\href{https://arxiv.org/abs/2108.09948}{{\ttfamily 2108.09948}}].

\bibitem{BESIII:2020iwk}
{\scshape BESIII} collaboration, \emph{{Search for the lepton number violating
  decay $\Sigma^{-} \to p e^{-} e^{-}$ and the rare inclusive decay $\Sigma^{-}
  \to \Sigma^{+} X$}},
  \href{https://doi.org/10.1103/PhysRevD.103.052011}{\emph{Phys. Rev. D}
  {\bfseries 103} (2021) 052011}
  [\href{https://arxiv.org/abs/2012.03592}{{\ttfamily 2012.03592}}].

\bibitem{LHCb:2017rdd}
{\scshape LHCb} collaboration, \emph{{Evidence for the rare decay $\Sigma^+ \to
  p \mu^+ \mu^-$}},
  \href{https://doi.org/10.1103/PhysRevLett.120.221803}{\emph{Phys. Rev. Lett.}
  {\bfseries 120} (2018) 221803}
  [\href{https://arxiv.org/abs/1712.08606}{{\ttfamily 1712.08606}}].

\bibitem{HyperCP:2005mvo}
{\scshape HyperCP} collaboration, \emph{{Evidence for the decay $\Sigma^+\to
  p\mu^+\mu^-$}},
  \href{https://doi.org/10.1103/PhysRevLett.94.021801}{\emph{Phys. Rev. Lett.}
  {\bfseries 94} (2005) 021801}
  [\href{https://arxiv.org/abs/hep-ex/0501014}{{\ttfamily hep-ex/0501014}}].

\bibitem{PiENu:2015seu}
{\scshape PiENu} collaboration, \emph{{Improved Measurement of the $\pi \to
  \textrm{e} \nu$ Branching Ratio}},
  \href{https://doi.org/10.1103/PhysRevLett.115.071801}{\emph{Phys. Rev. Lett.}
  {\bfseries 115} (2015) 071801}
  [\href{https://arxiv.org/abs/1506.05845}{{\ttfamily 1506.05845}}].

\bibitem{PIONEER:2022yag}
{\scshape PIONEER} collaboration, \emph{{PIONEER: Studies of Rare Pion
  Decays}},  \href{https://arxiv.org/abs/2203.01981}{{\ttfamily 2203.01981}}.

\bibitem{PIONEER:2022alm}
{\scshape PIONEER} collaboration, \emph{{Testing Lepton Flavor Universality and
  CKM Unitarity with Rare Pion Decays in the PIONEER experiment}},  in
  \emph{{2022 Snowmass Summer Study}}, 3, 2022
  [\href{https://arxiv.org/abs/2203.05505}{{\ttfamily 2203.05505}}].

\bibitem{Bryman:2011zz}
D.~Bryman, W.J.~Marciano, R.~Tschirhart and T.~Yamanaka, \emph{{Rare kaon and
  pion decays: Incisive probes for new physics beyond the standard model}},
  \href{https://doi.org/10.1146/annurev-nucl-102010-130431}{\emph{Ann. Rev.
  Nucl. Part. Sci.} {\bfseries 61} (2011) 331}.

\bibitem{Pocanic:2014jka}
D.~Pocanic, E.~Frlez and A.~van~der Schaaf, \emph{{Experimental study of rare
  charged pion decays}},
  \href{https://doi.org/10.1088/0954-3899/41/11/114002}{\emph{J. Phys. G}
  {\bfseries 41} (2014) 114002}
  [\href{https://arxiv.org/abs/1407.2865}{{\ttfamily 1407.2865}}].

\bibitem{PEN:2018kgj}
{\scshape PEN} collaboration, \emph{{PEN experiment: a precise test of lepton
  universality}},  in \emph{{13th Conference on the Intersections of Particle
  and Nuclear Physics}}, 11, 2018
  [\href{https://arxiv.org/abs/1812.00782}{{\ttfamily 1812.00782}}].

\bibitem{Mazza:2021adt}
{\scshape PIONEER} collaboration, \emph{{An LGAD-Based Full Active Target for
  the PIONEER Experiment}},
  \href{https://doi.org/10.3390/instruments5040040}{\emph{Instruments}
  {\bfseries 5} (2021) 40} [\href{https://arxiv.org/abs/2111.05375}{{\ttfamily
  2111.05375}}].

\bibitem{Mihara:2011zza}
S.~Mihara, \emph{{MEG liquid xenon detector}},
  \href{https://doi.org/10.1088/1742-6596/308/1/012009}{\emph{J. Phys. Conf.
  Ser.} {\bfseries 308} (2011) 012009}.

\bibitem{MEG:2016leq}
{\scshape MEG} collaboration, \emph{{Search for the lepton flavour violating
  decay $\mu ^+ \rightarrow \mathrm {e}^+ \gamma $ with the full dataset of the
  MEG experiment}},
  \href{https://doi.org/10.1140/epjc/s10052-016-4271-x}{\emph{Eur. Phys. J. C}
  {\bfseries 76} (2016) 434}
  [\href{https://arxiv.org/abs/1605.05081}{{\ttfamily 1605.05081}}].

\bibitem{MEGII:2018kmf}
{\scshape MEG II} collaboration, \emph{{The design of the MEG II experiment}},
  \href{https://doi.org/10.1140/epjc/s10052-018-5845-6}{\emph{Eur. Phys. J. C}
  {\bfseries 78} (2018) 380}
  [\href{https://arxiv.org/abs/1801.04688}{{\ttfamily 1801.04688}}].

\bibitem{PIENU:2019usb}
{\scshape PIENU} collaboration, \emph{{Search for heavy neutrinos in $\pi \to
  \mu\nu$ decay}},
  \href{https://doi.org/10.1016/j.physletb.2019.134980}{\emph{Phys. Lett. B}
  {\bfseries 798} (2019) 134980}
  [\href{https://arxiv.org/abs/1904.03269}{{\ttfamily 1904.03269}}].

\bibitem{Pocanic:2003pf}
D.~Pocanic et~al., \emph{{Precise measurement of the $\pi^+\to\pi^0 e^+\nu$
  branching ratio}},
  \href{https://doi.org/10.1103/PhysRevLett.93.181803}{\emph{Phys. Rev. Lett.}
  {\bfseries 93} (2004) 181803}
  [\href{https://arxiv.org/abs/hep-ex/0312030}{{\ttfamily hep-ex/0312030}}].

\bibitem{Feng:2020zdc}
X.~Feng, M.~Gorchtein, L.-C.~Jin, P.-X.~Ma and C.-Y.~Seng,
  \emph{{First-principles calculation of electroweak box diagrams from lattice
  QCD}}, \href{https://doi.org/10.1103/PhysRevLett.124.192002}{\emph{Phys. Rev.
  Lett.} {\bfseries 124} (2020) 192002}
  [\href{https://arxiv.org/abs/2003.09798}{{\ttfamily 2003.09798}}].

\bibitem{Czarnecki:2019iwz}
A.~Czarnecki, W.J.~Marciano and A.~Sirlin, \emph{{Pion beta decay and
  Cabibbo-Kobayashi-Maskawa unitarity}},
  \href{https://doi.org/10.1103/PhysRevD.101.091301}{\emph{Phys. Rev. D}
  {\bfseries 101} (2020) 091301}
  [\href{https://arxiv.org/abs/1911.04685}{{\ttfamily 1911.04685}}].

\bibitem{Gan:2020aco}
L.~Gan, B.~Kubis, E.~Passemar and S.~Tulin, \emph{{Precision tests of
  fundamental physics with $\eta$ and $\eta^\prime$ mesons}},
  \href{https://doi.org/10.1016/j.physrep.2021.11.001}{\emph{Phys. Rept.}
  {\bfseries 945} (2022) 2191}
  [\href{https://arxiv.org/abs/2007.00664}{{\ttfamily 2007.00664}}].

\bibitem{Feng:2016jff}
J.L.~Feng, B.~Fornal, I.~Galon, S.~Gardner, J.~Smolinsky, T.M.P.~Tait et~al.,
  \emph{{Protophobic Fifth-Force Interpretation of the Observed Anomaly in
  $^8$Be Nuclear Transitions}},
  \href{https://doi.org/10.1103/PhysRevLett.117.071803}{\emph{Phys. Rev. Lett.}
  {\bfseries 117} (2016) 071803}
  [\href{https://arxiv.org/abs/1604.07411}{{\ttfamily 1604.07411}}].

\bibitem{Nelson:1989fx}
A.E.~Nelson and N.~Tetradis, \emph{{CONSTRAINTS ON A NEW VECTOR BOSON COUPLED
  TO BARYONS}}, \href{https://doi.org/10.1016/0370-2693(89)90196-2}{\emph{Phys.
  Lett. B} {\bfseries 221} (1989) 80}.

\bibitem{Tulin:2014tya}
S.~Tulin, \emph{{New weakly-coupled forces hidden in low-energy QCD}},
  \href{https://doi.org/10.1103/PhysRevD.89.114008}{\emph{Phys. Rev. D}
  {\bfseries 89} (2014) 114008}
  [\href{https://arxiv.org/abs/1404.4370}{{\ttfamily 1404.4370}}].

\bibitem{Batell:2018fqo}
A.I.~B.~Batell, A.~Freitas and D.~McKeen, \emph{Probing light dark matter with
  a hadrophilic scalar mediator},
  \href{https://doi.org/doi:10.1103/PhysRevD.100.095020}{\emph{Phys. Rev. D}
  {\bfseries 100} (2019) 095020}.

\bibitem{Liu:2018qgl}
Y.-S.~Liu, I.C.~Clo\"et and G.A.~Miller, \emph{{Eta Decay and Muonic Puzzles}},
  \href{https://doi.org/10.1016/j.nuclphysb.2019.114638}{\emph{Nucl. Phys.}
  {\bfseries B} (2019) 114638}
  [\href{https://arxiv.org/abs/1805.01028}{{\ttfamily 1805.01028}}].

\bibitem{Bell:1968wta}
J.S.~Bell and D.G.~Sutherland, \emph{{Current algebra and $\eta\to 3\pi$}},
  \href{https://doi.org/10.1016/0550-3213(68)90316-7}{\emph{Nucl. Phys. B}
  {\bfseries 4} (1968) 315}.

\bibitem{Sutherland:1966zz}
D.G.~Sutherland, \emph{{Current algebra and the decay $\eta \to 3\pi$}},
  \href{https://doi.org/10.1016/0031-9163(66)90477-X}{\emph{Phys. Lett.}
  {\bfseries 23} (1966) 384}.

\bibitem{Baur:1995gc}
R.~Baur, J.~Kambor and D.~Wyler, \emph{{Electromagnetic corrections to the
  decays $\eta\to 3\pi$}},
  \href{https://doi.org/10.1016/0550-3213(95)00643-5}{\emph{Nucl. Phys. B}
  {\bfseries 460} (1996) 127}
  [\href{https://arxiv.org/abs/hep-ph/9510396}{{\ttfamily hep-ph/9510396}}].

\bibitem{Ditsche:2008cq}
C.~Ditsche, B.~Kubis and U.-G.~Meissner, \emph{{Electromagnetic corrections in
  $\eta\to 3\pi$ decays}},
  \href{https://doi.org/10.1140/epjc/s10052-009-0863-z}{\emph{Eur. Phys. J. C}
  {\bfseries 60} (2009) 83} [\href{https://arxiv.org/abs/0812.0344}{{\ttfamily
  0812.0344}}].

\bibitem{Colangelo:2018jxw}
G.~Colangelo, S.~Lanz, H.~Leutwyler and E.~Passemar, \emph{{Dispersive analysis
  of $\eta \rightarrow 3 \pi $}},
  \href{https://doi.org/10.1140/epjc/s10052-018-6377-9}{\emph{Eur. Phys. J. C}
  {\bfseries 78} (2018) 947}
  [\href{https://arxiv.org/abs/1807.11937}{{\ttfamily 1807.11937}}].

\bibitem{Schneider:2010hs}
S.P.~Schneider, B.~Kubis and C.~Ditsche, \emph{{Rescattering effects in
  $\eta\to 3\pi$ decays}},
  \href{https://doi.org/10.1007/JHEP02(2011)028}{\emph{JHEP} {\bfseries 02}
  (2011) 028} [\href{https://arxiv.org/abs/1010.3946}{{\ttfamily 1010.3946}}].

\bibitem{Colangelo:2022jxc}
G.~Colangelo et~al., \emph{{Prospects for precise predictions of $a_\mu$ in the
  Standard Model}},  \href{https://arxiv.org/abs/2203.15810}{{\ttfamily
  2203.15810}}.

\bibitem{Sanchez-Puertas:2018tnp}
P.~Sanchez-Puertas, \emph{{$CP$ violation in $\eta$ muonic decays}},
  \href{https://doi.org/10.1007/JHEP01(2019)031}{\emph{JHEP} {\bfseries 01}
  (2019) 031} [\href{https://arxiv.org/abs/1810.13228}{{\ttfamily
  1810.13228}}].

\bibitem{Gardner:2019nid}
S.~Gardner and J.~Shi, \emph{{Patterns of CP violation from mirror symmetry
  breaking in the $\eta\to\pi^+\pi^-\pi^0$ Dalitz plot}},
  \href{https://doi.org/10.1103/PhysRevD.101.115038}{\emph{Phys. Rev. D}
  {\bfseries 101} (2020) 115038}
  [\href{https://arxiv.org/abs/1903.11617}{{\ttfamily 1903.11617}}].

\bibitem{Shi:2020ffh}
J.~Shi, \emph{{Theoretical Studies of C and CP Violation in $\eta \to \pi^+
  \pi^- \pi^0$ Decay}}, Ph.D. thesis, Kentucky U., 2020.
\newblock 10.13023/etd.2020.388.

\bibitem{REDTOP:2022slw}
{\scshape REDTOP} collaboration, \emph{{The REDTOP experiment: Rare
  $\eta/\eta^{\prime}$ Decays To Probe New Physics}},
  \href{https://arxiv.org/abs/2203.07651}{{\ttfamily 2203.07651}}.

\bibitem{Akdag:2021efj}
H.~Akdag, T.~Isken and B.~Kubis, \emph{{Patterns of C- and CP-violation in
  hadronic $\eta$ and $\eta^\prime$ three-body decays}},
  \href{https://doi.org/10.1007/JHEP02(2022)137}{\emph{JHEP} {\bfseries 02}
  (2022) 137} [\href{https://arxiv.org/abs/2111.02417}{{\ttfamily
  2111.02417}}].

\bibitem{Jung:2010ik}
M.~Jung, A.~Pich and P.~Tuzon, \emph{{Charged-Higgs phenomenology in the
  Aligned two-Higgs-doublet model}},
  \href{https://doi.org/10.1007/JHEP11(2010)003}{\emph{JHEP} {\bfseries 11}
  (2010) 003} [\href{https://arxiv.org/abs/1006.0470}{{\ttfamily 1006.0470}}].

\bibitem{Becirevic:2016yqi}
D.~Be\v{c}irevi\'c, S.~Fajfer, N.~Ko\v{s}nik and O.~Sumensari,
  \emph{{Leptoquark model to explain the $B$-physics anomalies, $R_K$ and
  $R_D$}}, \href{https://doi.org/10.1103/PhysRevD.94.115021}{\emph{Phys. Rev.
  D} {\bfseries 94} (2016) 115021}
  [\href{https://arxiv.org/abs/1608.08501}{{\ttfamily 1608.08501}}].

\bibitem{Garces:2017jpz}
E.A.~Garc\'es, M.~Hern\'andez~Villanueva, G.~L\'opez~Castro and P.~Roig,
  \emph{{Effective-field theory analysis of the $\tau^- \to \eta^{(\prime)}
  \pi^- \nu_\tau$ decays}},
  \href{https://doi.org/10.1007/JHEP12(2017)027}{\emph{JHEP} {\bfseries 12}
  (2017) 027} [\href{https://arxiv.org/abs/1708.07802}{{\ttfamily
  1708.07802}}].

\bibitem{pro-JEF14}
{M. Dugger et al.}, ``{Eta Decays with Emphasis on Rare Neutral Modes: The JLab
  Eta Factory (JEF) Experiment}.'' 2014.

\bibitem{pro-JEF17}
{H. Al Ghoul et al.}, ``{Update to the JEF proposal (PR12-14-004)}.'' 2017.

\bibitem{GlueX:2020idb}
{\scshape GlueX} collaboration, \emph{{The GlueX Beamline and Detector}},
  \href{https://doi.org/10.1016/j.nima.2020.164807}{\emph{Nucl. Instrum. Meth.
  A} {\bfseries 987} (2021) 164807}
  [\href{https://arxiv.org/abs/2005.14272}{{\ttfamily 2005.14272}}].

\bibitem{Asaturyan:2021ese}
A.~Asaturyan et~al., \emph{{Electromagnetic calorimeters based on scintillating
  lead tungstate crystals for experiments at Jefferson Lab}},
  \href{https://doi.org/10.1016/j.nima.2021.165683}{\emph{Nucl. Instrum. Meth.
  A} {\bfseries 1013} (2021) 165683}.

\bibitem{Adlarson:2019nwa}
{\scshape A2} collaboration, \emph{{Recent Results on Meson Decays from A2}},
  \href{https://doi.org/10.1051/epjconf/201921803002}{\emph{EPJ Web Conf.}
  {\bfseries 218} (2019) 03002}.

\bibitem{Kashevarov:2017kqb}
{\scshape A2} collaboration, \emph{{Study of $\eta$ and $\eta^\prime$
  Photoproduction at MAMI}},
  \href{https://doi.org/10.1103/PhysRevLett.118.212001}{\emph{Phys. Rev. Lett.}
  {\bfseries 118} (2017) 212001}
  [\href{https://arxiv.org/abs/1701.04809}{{\ttfamily 1701.04809}}].

\bibitem{Husken:2019dou}
{\scshape WASA-at-COSY} collaboration, \emph{{$\eta$ meson physics with
  WASA-at-COSY}}, \href{https://doi.org/10.1051/epjconf/201919901006}{\emph{EPJ
  Web Conf.} {\bfseries 199} (2019) 01006}.

\bibitem{Berlowski:2019hst}
{\scshape KLOE-2} collaboration, \emph{{$\eta$ and $\eta^\prime$ physics in
  KLOE}}, \href{https://doi.org/10.1051/epjconf/201921808003}{\emph{EPJ Web
  Conf.} {\bfseries 218} (2019) 08003}.

\bibitem{Shuang-shiFang:2015vva}
{\scshape BESIII} collaboration, \emph{{$\eta$ and $\eta$' physics at BESIII}},
  \href{https://doi.org/10.22323/1.253.0032}{\emph{PoS} {\bfseries CD15} (2016)
  032}.

\end{thebibliography}\endgroup
